# Science goals and mission concepts for a future orbital and *in situ* exploration of Titan

White Paper submission to the call for ESA Voyage 2050 long-term plan

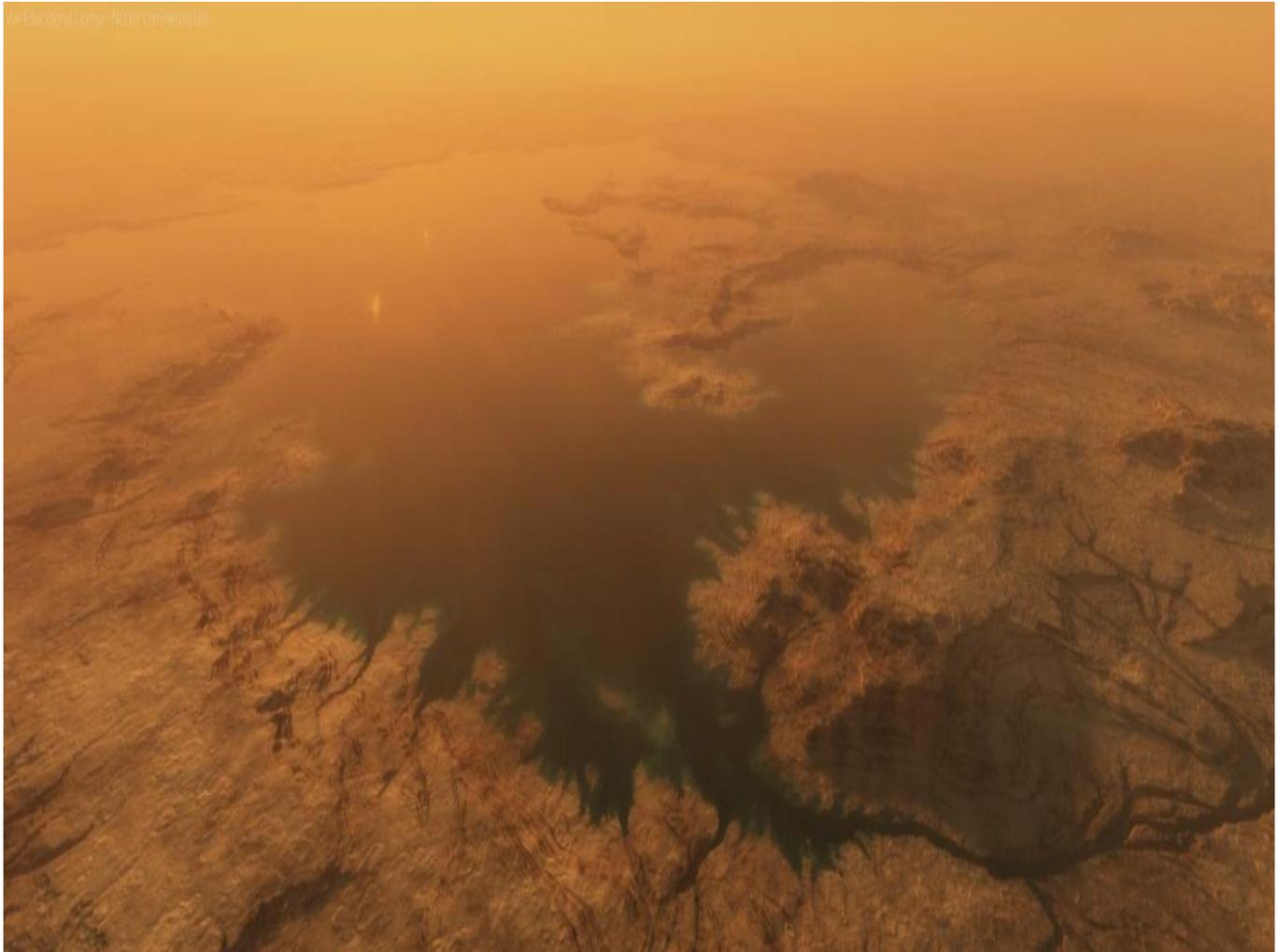

**Bird's-eye view of the hydrocarbon sea Ligeia Mare, North Pole of Titan. The topography is extracted from Cassini RADAR SAR images and textured using the same set of images. The view has been realistically colored and illuminated. (credits: Université de Paris/IPGP/CNRS/A. Lucas)**

**Contact Person:**
Sébastien Rodriguez
Planetary and Space Science group
Université de Paris, Institut de physique du globe de Paris, CNRS
Bâtiment Lamarck A- Bureau 723
35 rue Hélène Brion
75013 Paris, France
rodriguez@ipgp.fr
+33157275424
+33687002018

# Core Proposing Team

**Sébastien Rodriguez**,
Université de Paris, Institut de physique du globe de Paris, CNRS, F-75005 Paris, France

**Sandrine Vinatier**,
LESIA, Observatoire de Paris, Université PSL, CNRS, Sorbonne Université, Université de Paris, 5 place Jules Janssen, 92195 Meudon, France

**Daniel Cordier**,
Groupe de Spectrométrie Moléculaire et Atmosphérique, UMR CNRS 7331, Université de Reims Champagne-Ardenne, Reims, France

**Nathalie Carrasco**,
LATMOS, UMR CNRS 8190, Université Versailles St Quentin, Guyancourt, France. 3Institut Universitaire de France, Paris, France

**Benjamin Charnay**,
LESIA, Observatoire de Paris, Université PSL, CNRS, Sorbonne Université, Université de Paris, 5 place Jules Janssen, 92195 Meudon, France

**Thomas Cornet**,
European Space Agency (ESA), European Space Astronomy Centre (ESAC), Villanueva de la Canada, Madrid, Spain

**Athena Coustenis**,
LESIA, Observatoire de Paris, Université PSL, CNRS, Sorbonne Université, Université de Paris, 5 place Jules Janssen, 92195 Meudon, France

**Remco de Kok**,
Department of Physical Geography, Utrecht University, Utrecht, Netherlands

**Caroline Freissinet**,
LATMOS, UMR CNRS 8190, Université Versailles St Quentin, Guyancourt, France. 3Institut Universitaire de France, Paris, France

**Marina Galand**,
Department of Physics, Imperial College London, Prince Consort Road, London SW7 2AZ, UK

**Wolf D. Geppert**,
Department of Physics, AlbaNova University Center, Stockholm University, Roslagstullsbacken 21, Stockholm SE-10691, Sweden

**Ralf Jauman**,
DLR, Institute of Planetary Research, Berlin, Germany

**Klara Kalousova**,
Charles University, Faculty of Mathematics and Physics, Department of Geophysics, Prague, Czech Republic

**Tommi T. Koskinen**,
Lunar and Planetary Laboratory, University of Arizona, 1629 E. University Blvd., Tucson, AZ 85721, USA

**Sébastien Lebonnois**,
Laboratoire de Météorologie Dynamique (LMD/IPSL), Sorbonne Université, ENS, PSL Research University, Ecole Polytechnique, Université Paris Saclay, CNRS, Paris, France

**Alice Le Gall**,
LATMOS, UMR CNRS 8190, Université Versailles St Quentin, Guyancourt, France. 3Institut Universitaire de France, Paris, France

**Stéphane Le Mouélic**,
Laboratoire de Planétologie et Géodynamique, UMR 6112, CNRS, Université de Nantes, 2 rue de la Houssinière, Nantes 44322, France

**Antoine Lucas**,
Université de Paris, Institut de physique du globe de Paris, CNRS, F-75005 Paris, France

**Kathleen Mandt**,
Johns Hopkins University Applied Physics Laboratory, 11100 Johns Hopkins Rd., Laurel, MD 20723, USA

**Marco Mastrogiuseppe**,
Division of Geological and Planetary Sciences, California Institute of Technology, Pasadena, CA, USA

**Conor A. Nixon**,
Planetary Systems Laboratory, NASA Goddard Space Flight Center, Greenbelt, MD 20771, USA

**Jani Radebaugh**,
Department of Geological Sciences, Brigham Young University, S-389 ESC Provo, UT 84602, United States

**Pascal Rannou**,
Groupe de Spectrométrie Moléculaire et Atmosphérique, UMR CNRS 7331, Université de Reims Champagne-Ardenne, Reims, France

**Jason M. Soderblom**,
Department of Earth, Atmospheric and Planetary Sciences, Massachusetts Institute of Technology, Cambridge, Massachusetts, USA

**Anezina Solomonidou**,
European Space Agency (ESA), European Space Astronomy Centre (ESAC), Villanueva de la Canada, Madrid, Spain

**Christophe Sotin**,
Jet Propulsion Laboratory, Caltech, 4800 Oak Grove Drive, Pasadena, CA 91109, USA

**Katrin Stephan**,
DLR, Institute of Planetary Research, Berlin, Germany

**Nick Teanby**,
School of Earth Sciences, University of Bristol, Wills Memorial Building, Queens Road, Bristol, BS8 1RJ, UK

**Gabriel Tobie**,
Laboratoire de Planétologie et Géodynamique, UMR 6112, CNRS, Université de Nantes, 2 rue de la Houssinière, Nantes 44322, France

**Véronique Vuitton**,
Institut de Planétologie et d'Astrophysique de Grenoble, Univ. Grenoble Alpes, CNRS, Grenoble 38000, France

# 1. Context, motivations and new mission concept summary

Saturn's giant moon, Titan, is one of the Solar System's most enigmatic bodies and is therefore a prime target for future space exploration, all the more so after the Cassini-Huygens mission, which has demonstrated that Titan provides an analogue for many processes relevant to Earth, outer Solar System bodies, and a growing number of newly discovered exoplanets. Important and quite unique processes are happening on this body, including complex organic chemistry of astrobiological relevance, methane meteorological cycles, surface liquids and lakes, geology, fluvial and aeolian erosion, and unique interactions of the atmosphere with an external plasma environment.

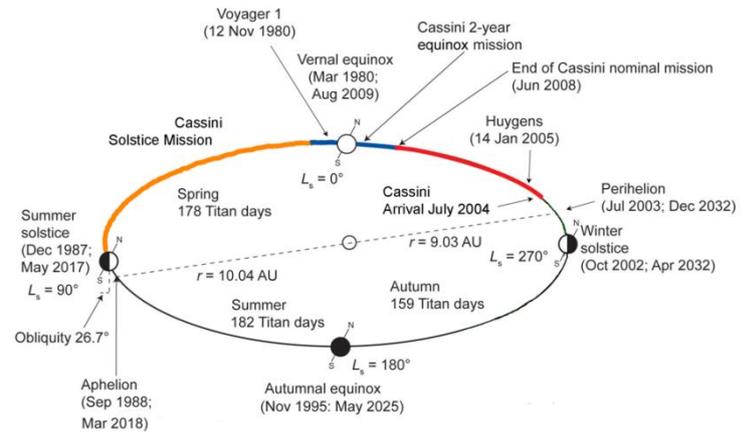

*Figure 1: Titan experience long seasons of about 7.5 Earth years. A mission arriving in 2030-2040s will encounter similar seasons to Cassini-Huygens allowing for data set inter-comparison and evaluation of interannual changes. This will also provide an excellent complement to the newly selected Dragonfly mission that should arrive at Titan in 2034 (from "Titan Explorer: Flagship Mission Study" by J. C. Leary et al., Jan 2008).*

The entry of the Cassini-Huygens spacecraft into orbit around Saturn in July 2004 marked the start of a golden era in the exploration of Titan. During the prime mission (2004-2008), ground-breaking discoveries were made by the Cassini orbiter including the presence of equatorial dune fields, northern lakes and seas, large positive and negative ions in the upper atmosphere. In 2005 the Huygens probe descended through Titan's atmosphere, taking the first close-up pictures of the surface, revealing large networks of dendritic channels leading to a dried-up seabed, and also obtaining detailed profiles of temperature and gas composition of the atmosphere. The discoveries continued through the Equinox mission (2008-2010) and Solstice mission (2010-2017) performing a total of 127 targeted flybys of Titan. Now, after the end of the mission, we are able to look back on the high level scientific questions from the beginning of the mission, and assess the progress that has been made towards their answering.

At the same time, new important scientific questions regarding Titan have emerged from the discoveries that have been made to date. Cassini was a dedicated mission to study the entire Saturn system and the limited number of Titan's flybys did not have a sufficient frequency (in average one flyby per month) to monitor atmospheric processes varying within a few hours/days (like variability in the thermosphere or clouds in the deep atmosphere, **Sections 2.1 and 2.3**) and to totally apprehend the complexity of its surface dynamics and interior structure (aeolian and fluid transport, formation and evolution of lakes and seas, erosion, cryovolcanic activity, depth and thickness of its ice shell and global ocean…, **Section 3**). Regarding the atmosphere, the region between 500 and 1000 km, informally called the "agnostosphere", remains poorly known, as it was only probed *in situ* by Huygens (at a single location and time) and could only be studied using the few solar/stellar occultations measurements by the UV spectrometer (UVIS, Esposito et al. 2004). Cassini *in situ* measurements above 1000 km altitude revealed a complex chemistry in the upper atmosphere (see **Section 2.1**), but only the smallest ionic species (< 100 amu) could be identified by the Ion and Neutral Mass spectrometer (INMS, Waite et al. 2004), whereas the presence of large amounts of anions with higher masses up to 10,000 amu/q were detected by the Cassini Plasma Spectrometer (CAPS, Young et al. 2004). Further, Cassini only indirectly constrained the global atmospheric circulation in the lowest altitude regions below 500 km (**Sections 2.2 and 2.3**) (with the exception of the wind speed measurements of the Huygens probe during its descent near the equator). Many aspects of the complexity of the climatology of the moon in the lowest part of its atmosphere are also far from being fully understood, such as the distribution and seasonality of cloud formation, the intensity of methane evaporation and precipitation, the origin and impact of atmospheric waves, or the intensity and direction of surface winds (**Section 2.3**). In the same manner, important questions remain regarding Titan's surface and interior. To name a few, the age of Titan's surface is still poorly constrained (**Section 3.4**), the depth and thickness of its ice shell and global ocean are still unknown and past or present cryovolcanic activity has still not been detected (**Section 3.5**), its absolute surface reflectivity and composition are almost completely unknown (relevant to all topics of **Section 3**), the origin and morphodynamics of dunes, rivers, lakes and seas, and associated erosion rates, are still strongly debated (**Sections 3.1, 3.2 and 3.3**). Finally, following the established definition for habitability (presence of a stable substrate, available energy, organic chemistry, and the potential for holding a liquid solvent), Titan is one of celestial bodies in the solar system with the highest potential for habitability. If Cassini-Huygens brought essential observations to sustain this hypothesis, the real habitability potential of Titan is still strongly debated (what source(s) of energy and where?, complexity of the organic chemistry?, which solvent(s)?, **Section 4**). We will not have more information on these fundamental questions without a new and ambitious exploration program specially dedicated to Titan.

In this white paper, we present a cross-section of important scientific questions that remain partially or completely unanswered, ranging from Titan's exosphere to the deep interior, and we detail which instrumentation and mission scenarios should be used to answer them. Our intention is to formulate the science goals for the next generation of planetary missions to Titan in order to prepare the future exploration of the moon. The ESA L-class mission concept that we propose is composed of a Titan orbiter and at least an *in situ* element (lake lander and /or drone(s)). The choice of an orbiter would guarantee global coverage, good repetitiveness and high spatial resolution for Titan's atmosphere and surface observations both in imagery and spectroscopy. Ideally the orbiter would have a high inclination elliptical orbit whose closest approach altitude will be localized in the thermosphere to perform *in situ* mass spectrometry measurements at each orbit. A – preferably mobile – *in situ* package (drone(s)) would be dedicated to study Titan's areas of particular geological interest at unprecedented coverage and spatial resolution. It would also be used to perform measurements in the atmosphere during the descent. The *in situ* element(s) should preferentially be sent at high northern latitudes to study the lake/sea region and to perform atmospheric measurements inside the polar vortex. The ideal arrival time at Titan would be slightly before 2039 (the next Northern Spring equinox, while the Northern Fall equinox will not occur before 2054), as equinoxes are the most exciting observing periods to monitor the most striking and still largely unknown atmospheric and surface seasonal changes. In addition, a mission arriving in 2030s will encounter similar seasons to Cassini-Huygens allowing for data set inter-comparison and evaluation of interannual changes (**Figure 1**). With an arrival slightly before 2039, the presence of an orbiter and the focused exploration of Titan's Northern latitudes would provide an extraordinary complementary in terms of timing (with possible mission timing overlap), locations and science goals with the Dragonfly mission. In the case of a partnership of ESA with NASA regarding the Dragonfly mission, our arrival at Titan should be as early as 2034. Note that any arrival outside Dragonfly mission calendar, or equinoxes, would still have, along with the use of an orbiter and an *in situ* element, an outstanding scientific value, still answering fundamental open questions that remain about Titan's system that cannot be answered from Earth ground-based/space-borne facilities.

## 2. Science Goal A: Titan's atmosphere

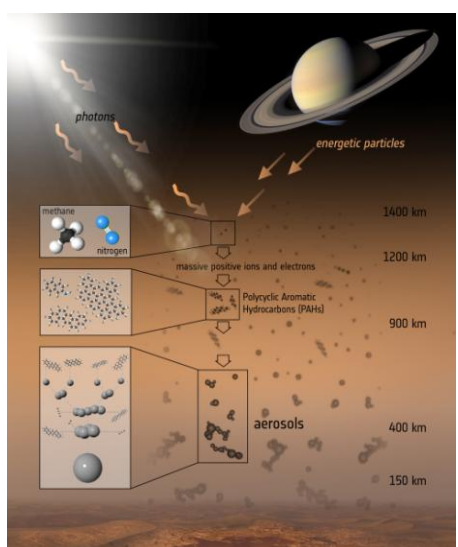

Figure 2: Titan's atmosphere consists of the two primary gases, methane ($CH_4$) and nitrogen ($N_2$) that undergo a series of photochemical reactions to produce heavier molecules, and the ubiquitous haze particles (aerosols). (ESA/ATG medialab)

Titan's atmosphere shows the most complex chemistry in the solar system. Micrometer-size aerosols are formed from a chain of chemical reactions initiated in the upper atmosphere by ionization and dissociation of nitrogen and methane by solar UV photons and associated photoelectrons (Galand et al. 2010) (**Figure 2**). The photochemically produced molecules and aerosols have a strong impact on the radiative budget of Titan's atmosphere, and consequently on its climate. Besides, transport by global dynamics, which reverses each half Titan-year, greatly affects the distribution of these compounds. The complex couplings of chemistry, radiation, and dynamics make Titan's atmosphere an ideal laboratory to understand physical and chemical processes at play in atmospheres and particularly those showing the presence of photochemical haze such as Pluto (Cheng et al. 2017) or the increasing number of exoplanets (Pinhas et al. 2019). Furthermore, the study of Titan's chemistry has astrobiological implications as it naturally produces nitrogen-containing organic molecules, which can act as biomolecule precursors. Titan's atmosphere also contains oxygen compounds, which could play a role in the formation of amino-acids and DNA nucleobases, as suggested by laboratory experiments that simulate Titan's chemistry (Hörst et al. 2012).

**2.1 The upper atmosphere: a region of complex physical and chemical processes**

*2.1.1 Current knowledge*

One of the most outstanding observations of the Cassini mission was the in-situ detection of ions with several hundred mass units above 900 km (Coates et al. 2007; Waite et al. 2007) revealing the dusty plasma nature of Titan's upper atmosphere (located above 550 km altitude) and its highly complex ion-neutral chemistry (Vuitton et al. 2006, 2007) producing the nanoparticles at much higher altitude than previously thought (Lavvas et al. 2013). On Titan's dayside, the main source of ionization is the solar extreme ultraviolet (EUV) radiation while ionization from energetic electrons from Saturn's magnetosphere contributes as a source of night side short-lived ions (Cui et al., 2009; Sagnières et al. 2015; Vigren et al. 2015). In addition, day-to-night transport from the neutral atmosphere seems to be a significant source of the night side long-lived ions (Cui et al. 2009). Titan's thermosphere shows an unexpectedly variable $N_2$ (Titan's main gaseous species) density that changes by more than an order of magnitude on relatively short timescales (comparable to a few or less Titan's days), accompanied by large wave-like perturbations in temperature profiles (Westlake et al. 2011; Snowden et al. 2013). Titan, such as Venus, possesses an atmosphere in global superrotation, which extends up to at least 1000 km with a surprisingly high wind speed of 350 m/s (Lellouch et al. 2019). At altitudes of 500-1000 km, aerosol seed particles grow through coagulation and chemistry (Lavvas et al. 2011) while they drift downward into the deep atmosphere. Increasingly complex molecules are also produced by neutral photochemistry (Vuitton et al. 2019) and Cassini measurements, albeit limited in scope, revealed the density profiles of several key hydrocarbons and nitriles, the extinction due to aerosols and the possible presence of polycyclic aromatic hydrocarbons (PAHs) (Liang et al. 2007; Koskinen et al. 2011; Lopez-Puertas et al. 2011; Maltagliati et al. 2015).

*2.1.2 Open questions*

Cassini only partially revealed the role of ion-neutral chemistry in the upper atmosphere and could not address the chemical nature of heavy neutrals and ions and their formation mechanisms, nor the identity of macromolecules because of the limited mass range and resolution of Cassini *in situ* mass spectrometers (INMS had a mass resolution of m/Δm<500 at m/z=50). However, ion-neutral chemistry may dominate the formation of larger molecules above 900 km (Lavvas et al. 2011a). More generally, Titan's upper atmosphere offers the perfect natural laboratory to assess complex ion-neutral chemistry and to characterize dusty plasmas which are encountered in other astrophysics environments, such as the plumes of Enceladus (Morooka et al. 2011; Hill et al. 2012), and which are expected to be present in exoplanetary atmospheres.

Titan's atmosphere contains oxygen compounds (CO, $H_2O$, $CO_2$) and $O^+$ ions possibly sourced from Enceladus (Hörst et al. 2008; Dobrijevic et al. 2014). Whether oxygen is incorporated into the more complex organic molecules on Titan remains an outstanding question with great implications for prebiotic chemistry. If oxygen is detected in organics in the upper atmosphere, it can imply the presence of a global flux of prebiotic molecules descending to the surface.

The recent detection of strong thermospheric equatorial zonal winds with speeds increasing with height up to 350 m/s at 1000 km (Lellouch et al., 2019) was totally unexpected. The source of such rapid winds could be related to waves produced in the stratosphere in response to the diurnal variation of the solar insolation and propagating toward the upper atmosphere. These gravity waves, which were observed by Cassini-Huygens and in stellar occultations observed from the ground, could transfer momentum from the deepest atmospheric layers towards the upper atmosphere and accelerate equatorial winds. Monitoring the winds with latitude, local time, and season will give a clue to its origin. This will also be critical in order to assess the effect of high neutral winds on the ion densities, especially on the night side where transport from dayside is essential for long-lived ions (Cui et al. 2009).

A global picture of how the upper atmosphere works could not be derived from the Cassini measurements alone because the flybys were too scarce to monitor the density and temperature variability that seems to occur on timescales smaller than one Titan's day. No correlations between the temperature, and the latitudes, longitudes and solar insolation were found, which suggests that the temperature in this region is not controlled by the UV solar flux absorption. Understanding the origin of this variability is important because it affects the escape rates from Titan's atmosphere (Cui et al. 2011, 2012) and is likely to have consequences for the circulation and ion-neutral chemistry in the upper atmosphere.

To summarize, the main important questions related to the upper atmosphere are:
- **[1] What is the role of ion-neutral chemistry in the upper atmosphere?**
- **[2] What is the degree of oxygen incorporation in photochemical species?**
- **[3] What is the dynamics of the upper atmosphere and what is its origin? What is the effect on ion densities?**
- **[4] What are the physical processes in the ionosphere and thermosphere that drive their density and temperature variability?**

*2.1.3 Proposed instrumentation and mission concept to address the open questions*

**Understanding the upper atmospheric chemistry (open questions 1 & 2)** requires the identification of common subunits and building blocks that form the nanoparticles so that we can predict the chemical structures and reactivity of the large molecules. **This can only be done by using *in situ* analysis**. An ion and neutral mass spectrometer with a much higher mass resolution and a higher mass upper limit than INMS and CAPS aboard Cassini is absolutely required on a Titan orbiter. Such an instrument, like **Cosmorbitrap** (Briois et al. 2016) under development, will determine the composition of macromolecules up to 1000 u with a mass resolution m/Δm = 100,000 at m/z=100 (**Figure 4**) and a sensitivity of $10^{-3}$ molecule $cm^{-3}$.

**Understanding the variability of the thermosphere and the origin of supersonic winds (open questions 3 & 4)** in the upper atmosphere and how they evolve with season requires a **Titan's orbiter** carrying:
- **A submillimeter (submm) spectrometer,** in order to directly measure the wind from the Doppler shifts of the numerous spectral lines from the surface up to at least 1200 km with a vertical resolution of 8 km (Lellouch et al. 2010). This will provide the first 3D wind measurement in a celestial body other than the Earth. Such an instrument is also of prime interest to retrieving the thermal profile (from CO and HCN lines) and the mixing ratio profiles of many molecules ($H_2O$, $NH_3$, $CH_3C_2H$, $CH_2NH$, $HC_3N$, $HC_5N$, $CH_3CN$, $C_2H_3CN$, $C_2H_5CN$) and their isotopes in the same altitude range, giving crucial new insights in the chemistry of the upper atmosphere. These observations would be a major advance compared to what was done from Cassini, which mostly probed the 100-500 km region (from mid- and far-IR CIRS observations) and the ≈1000 km region (*in situ* with INMS and CAPS), while only a few UV solar/stellar occultations probed the 500-1000 km region. Here, it should be noted that ALMA will never provide a spatial resolution better than ≈400 km on Titan's disk.
- **An ultraviolet (UV) spectrometer,** necessary to reveal the atmospheric density variability (from $N_2$), the trace gas density variability ($CH_4$, $C_2H_2$, $C_2H_4$, $C_4H_2$, $C_6H_6$, HCN and $HC_3N$) and the aerosol extinction in the 500-1400 km range from stellar and solar occultations observed by a Titan orbiter. It will also measure the temperature structure at 500-1400 km. Observations of UV airglow and reflected light will constrain energy deposition and aerosol properties. To overcome the limitations of Cassini observations, the measurements should be designed to achieve high vertical resolution and targeted to map spatial and temporal trends in dedicated campaigns. Compared to Cassini/UVIS, the wavelength coverage should be extended to include the middle UV (MUV) range in addition to the EUV and FUV ranges. This would allow for better spectral constraints on aerosol extinction and possible detection of complex molecules below 1000 km that were not accessible to Cassini.

**Understanding physical processes in ionosphere (open questions 1, 3 & 4)** requires:
- **A pressure gauge,** providing absolute measurement of the total neutral density; such measurements would be combined with the neutral composition measurements from Cosmorbitrap (see above) to provide reliable absolute neutral densities. Cosmorbitrap would also provide critical positive ion composition measurements to identify the source of long-lived ions and the role played by transport through the thermospheric winds.
- **A Mutual Impedance Probe (MIP) (combined with a Langmuir Probe),** which provides absolute electron densities to be compared with the total positive ion densities from the positive ion mass spectrometer to highlight any difference between both in the region of dusty plasma. It would also provide electron temperatures relevant for assessing the energy budget and constraining ion-electron reaction coefficients for assessing ionospheric density.
- **A Negative Ion Mass Spectrometer,** which measures the composition and densities of negative ions (which had not been anticipated ahead of the Cassini mission, resulting of CAPS/ELS not being calibrated for such ions and not having suitable mass resolution) extending towards very high masses (as CAPS/ELS detected negative ions beyond 10,000 u/q (Coates et al. 2007)). This will provide critically new insights on the negative ion chemistry and formation of very negative ions (Vuitton et al. 2009).

In addition to these new sensors not yet flown at Titan, we would need complementary observations to provide context and critical observations for multi-instrument studies in order to address the raised questions:
- **A Langmuir probe,** measuring variability in electron density and temperature with high time resolution. It would also measure the spacecraft potential critical for interpreting ion and electron spectrometer dataset.
- **An Electron Spectrometer** (1 eV-1 keV), which measures the electron densities as a function of energy. This is essential for assessing the energetic electron population ionizing the upper atmosphere and assessing the relative importance of local ionization versus transport from dayside for the source of ions on the night side. It would also be essential for constraining the photoelectron source on the dayside.
- **A Magnetometer,** which measures the magnetic field vector components and, combined with 3D magnetospheric model, to derive the 3D configuration of the magnetic field lines, along which particles are transported. This is essential information for assessing the electron energy budget (Galand et al. 2006) and the interaction of Titan with Saturn's magnetosphere (Snowden et al. 2013). Magnetic field measurements would also be used for deriving pitch angle information for the electrons measured by the electron spectrometer.

**Understanding the upper atmospheric chemistry (open questions 3 & 4)** requires the identification of common subunits and building blocks composing the nanoparticles so we predict the chemical structures and reactivity of the large molecules. **This can only be done by using *in situ* analysis**. An ion and neutral mass spectrometer like the Cosmorbitrap (Briois et al. 2015) with a much higher mass resolution and a higher mass upper limit than INMS and CAPS aboard Cassini is absolutely required on a Titan orbiter.

It is important to emphasize that all the above required measurements can only be performed with a dedicated Titan orbiter as the autumn/winter polar region is not observable from the Earth and none of the largest current or future ground-based (ALMA, ELT) or space-borne facilities (JWST) will have enough spatial resolution to address the questions mentioned above.

**2.2 The middle atmosphere: global dynamics and its coupling to composition and haze distributions**

*2.2.1 Current knowledge*

The middle atmosphere is typically located above 80 km, where most molecules condense (with the exception of CO, $H_2$ and $C_2H_4$) at Titan's average temperature and pressure conditions. This region is a transition between the troposphere (**Section 2.3**), in which convection controls surface-atmosphere interactions, and the upper atmosphere, where molecules and aerosols are formed. Photochemical species' sources in the upper atmosphere and condensation sink in the deeper atmosphere result in increasing-with-height concentrations profiles, with vertical gradients strongly impacted by the global dynamics. This region contains the main haze layer with a permanent but altitude-variable detached haze layer on top during summer/winter. A significant correlation is observed between the detached haze layer position and a sharp transition in the temperature profile, marking the end of the mesosphere (and the boundary with the thermosphere) and the presence of large-scale gravity waves above (e.g. Porco et al. 2005; Fulchignoni et al. 2005). Like on Earth, a polar vortex forms on Titan during winter. Stratospheric polar vortices are regions of particular interest in planetary atmospheres. They are dominant dynamical structures, in which the air is isolated from the rest of the atmosphere by high speed zonal winds and whose morphology strongly varies with season. On Titan, a strong enrichment of photochemically produced species is observed inside polar vortices (e.g. Vinatier et al. 2015; Teanby et al. 2017; Coustenis et al. 2018), and massive polar stratospheric clouds with complex compositions have been detected during the northern winter and spring and during southern autumn (de Kok et al 2014; West et al. 2016; Le Mouelic et al. 2018). Atmospheric superrotation is intimately linked to the meridional circulation that shows marked seasonal changes with global pole-to-pole circulation during winter/summer and equator-to-pole circulation close to the equinoxes (every ≈15 years). This meridional circulation transports photochemical species (haze and molecules) that impact the radiative budget of the atmosphere and in turn affect the global dynamics. Aerosols especially strongly impact the radiative balance as they control the stratospheric temperature by diabatic heating in the visible and by dominating the cooling to space in the infrared especially during the winter polar night (Rannou et al., 2004; Larson et al., 2015; Bézard et al. 2018). Large-scale aerosol structures result from interaction with the atmospheric circulation, such as the global thin detached haze layer whose altitude drastically changed with season (West et al. 2011, 2018). A sharp minimum of zonal wind around 70-80 km altitude was observed in the lower stratosphere from Huygens *in situ* measurements (Bird et al. 2005) and indirectly derived from radio-occultation measurements (Flasar et al. 2013). This almost zero-wind layer decouples the tropospheric and stratospheric global dynamics.

Chemistry is also active in the middle atmosphere, while being less productive than in the upper atmosphere. Cassini/Composite InfraRed Spectrometer (CIRS, Flasar et al. 2004) and Visual and Infrared Mapping Spectrometer (VIMS, Brown et al. 2004) measurements provided a seasonal monitoring of the vertical and spatial distributions of only a dozen species (e.g. Vinatier et al. 2010; Coustenis et al. 2013; Teanby et al. 2019), constraining both 1-D photochemical models (e.g. Vuitton et al. 2019) and General Circulation Models (GCMs) (e.g. Lebonnois et al. 2012). The inventory of complex molecules was recently extended with the detection of $C_2H_5CN$ and $C_2H_3CN$ with ALMA (Cordiner et al. 2015, 2019; Palmer et al. 2017), albeit with limited spatial and vertical coverage. Those molecules could not be observed by Cassini due to the limited spectral coverage and/or sensitivity of the instruments.

*2.2.2 Open questions*

GCMs currently favor scenarios involving planetary-scale barotropic wave activity in the winter hemisphere (Read and Lebonnois 2018) to generate superrotation but these models have difficulties to maintain it. Possible signatures of these waves were detected recently (Cordier et al. private communication) on the haze spatial distribution from Cassini/Imaging Science Subsystem (ISS, Porco et al. 2004) images, but because of the limited number of flybys, only a few snapshots of their spatial and temporal evolutions, insufficient to constrain the models, are available. It is necessary to know how these waves evolve spatially on timescales from days to seasons to understand their impact on angular momentum transport and their role on generating and maintaining superrotation. This will be crucial to elucidate mechanisms at play in Titan's atmosphere and more generally for other partially superrotating atmospheres in the solar system or for tidally-locked exoplanets (e.g. Pierrehumbert 2011). The minimum zonal wind near 80 km altitude is currently not reproduced by GCMs and its consequences on angular momentum exchanges and transport of haze and trace species between troposphere and stratosphere are unknown.

During the Cassini mission, only a partial view of the vortex formation and its seasonal evolution was obtained, using (i) temperature and trace gas concentration distributions derived from CIRS (e.g. Achterberg et al 2011; Vinatier et al. 2015; Teanby et al. 2017; Coustenis et al. 2018), (ii) seasonal evolution of massive stratospheric polar clouds first observed at the north pole during northern winter and later at the south pole during southern autumn (Jennings et al. 2012; West et al. 2016; Le Mouelic et al. 2018). Polar vortex structures change on quite rapid timescales: for instance, the southern polar vortex has doubled in size within a few Titan's days in early autumn. We currently do not know what controls the latitudinal extent of the polar vortex and how it forms and disappears. Its vertical structure, while it was forming at the south pole, cannot be inferred between 2012 and 2015 because of too few Cassini limb observations of polar regions. Polar vortices are regions of strong interaction with the upper atmosphere as the subsiding air comes from above but the vortex structure across the upper atmosphere and stratosphere cannot be extracted from the Cassini observations as no UV occultations occurred in this region during the vortex formation phase.

The chemistry inside Titan's vortices is particularly interesting, showing the strongest enrichments in photochemically produced species and thus potentially producing even more complex chemistry than elsewhere in the middle atmosphere. The neutral atmosphere exhibits rapid and dramatic changes with seasons. The highest degree in complexity of the organic chemistry is unknown. The high molecular concentrations combined with the low temperatures result in condensation of almost all molecules in the lower stratosphere as high as 300 km altitude as observed at the south pole during autumn and result in the production of the most complex stratospheric ice clouds observed in the solar system. Composition of these massive clouds is currently poorly known. Only a few condensates have been identified: $HC_3N$ ice (Anderson et al. 2010), $C_4N_2$ ice (Anderson et al. 2016), HCN ice (de Kok et al. 2014; Le Mouélic et al. 2018), $C_6H_6$ ice (Vinatier et al. 2018) and co-condensed $HCN:C_6H_6$ ices (Anderson et al. 2018). These ice crystals then precipitate toward the polar lakes in which their chemical and potential astrobiological impacts are totally unknown.

Knowing the composition of Titan's aerosols is important to understand the chemistry occurring in the neutral atmosphere and to better characterize their impact on the radiative budget. Their composition is directly linked to the spectral variation of their refractive index, while their morphology (size, shape) affects their absorption and scattering properties. Aerosol refractive indices were roughly determined from Cassini/VIMS and CIRS observations in the 100-500 km range (Rannou et al. 2010; Vinatier et al. 2012) and haze optical properties were determined *in situ* at a single location below 150 km near the equator from Huygens' measurements of their scattering properties (Tomasko et al. 2008a; Doose et al. 2016). Knowledge of haze optical properties is also of prime importance to: (i) retrieve the surface spectra through the atmosphere in $CH_4$ windows in which aerosols have spectral contribution, especially at 2.8 μm near the strong N-H absorption peak; (ii) to understand the cloud formation as aerosol composition and their morphology influence the gases condensation. Titan's aerosols are also probably quite representative of the haze that seems to be naturally produced in $CH_4$-rich atmospheres like those of Pluto, Triton and probably many exoplanets.

Hence, many open questions cannot currently be answered without a dedicated mission to Titan:
- **[1] What generates Titan's atmospheric superrotation and what maintains it?**
- **[2] How do the polar vortices form, evolve and end?**
- **[3] What is the chemistry and its highest complexity attained inside polar vortices? What are the composition and the structure of the massive stratospheric polar clouds?**
- **[4] What are the composition and optical properties of aerosol in the main haze layer?**

*2.2.3 Proposed instrumentation and mission concept to address the open questions*

**Understanding the global dynamics of Titan's atmosphere (open question 1)** requires:
- **An orbiter visible and near IR imager,** monitoring in detail the wave activity around the equinox and when the polar vortex is forming. Monitoring the haze distribution using an orbiter imager will be the only way to constrain Titan's wave activity, as they are not detectable from Earth's orbit even from the largest facilities (JWST, ELT).
- **An orbiter submm spectrometer** to measure the wind profiles (including the 80 km altitude wind minimum) and probe the pole-to-pole structure along seasons. It will be possible to measure the meridional circulation speed for the first time (predicted to be at most of ≈1 m/s) by integrating ≈10 minutes with this type of instrument, which has a precision of 3 m/s in 1 min integration (Lellouch et al. 2010).
- **An orbiter radio occultation experiment** to probe the pole-to-pole structure of the 80 km altitude wind minimum along seasons as performed by Flasar et al. (2013) from the very limited number of radio occultations performed by Cassini.
- **An *in situ* wind measurement experiment** during the descent of one/several landers to get the precise profile and the changing directions of the winds along the descent.

**Understanding the polar vortex structure and its evolution (open question 2)** requires an orbiter with a **submm spectrometer** to directly measure the vortex zonal winds from a high inclination orbit (a strictly polar orbit would not permit however to measure zonal winds at high latitudes, Lellouch et al. 2010). Their spatial/vertical structures with time will be determined for the first time from the lower stratosphere to 1000 km. This will give the first 3D view of the vortex winds as well as its thermal structure and its composition in the middle and upper atmosphere where the vortex structure is totally unknown (as Cassini only probed altitudes below 600 km from thermal infrared measurements).

**The chemistry inside polar vortices (open question 3) and aerosol composition and optical properties (open question 4)** will be revealed with the combination of:
- **An orbiter submm spectrometer** providing the mixing ratios of many photochemical species from the stratosphere up to 1000 km.
- **An orbiter UV spectrometer** whose observations of stellar/solar occultations will probe the aerosol vertical structure from the 400-1400 km range as well as the density profiles of $CH_4$, HCN, $HC_3N$, and species that cannot be observed in the submm spectral range: $N_2$, $C_2H_2$, $C_2H_4$, $C_4H_2$ and $C_6H_6$.
- **An orbiter visible-near IR spectro-imager** necessary to determine from the reflected sunlight radiation, the optical constants of aerosol in the visible and near IR spectral range, and how it varies in the atmosphere and with season specially inside the winter polar vortex in which enriched air coming from above can modify their composition. This instrument will also probe the vertical profile of the aerosol extinction coefficient and will reveal the structure and composition of the clouds.
- **An orbiter far- to mid-IR spectrometer** necessary to determine the vertical and spatial variations of the aerosols optical constants in the far- and mid-IR spectral range where they emit IR radiation, and if these properties seasonally and spatially vary in the atmosphere. It will also allow the determination of $C_2H_2$, $C_2H_6$, $C_2H_4$, $C_4H_2$, $C_6H_6$ vertical profiles below 500 km, which cannot be observed in submm (no spectral lines) or UV (sensitive to altitudes higher than 450 km). This instrument should include the spectral range 1450 – 1900 cm$^{-1}$ (7 to 5 µm) that was not observed by Cassini. This range is of particular interest because it displays a strong peak of the haze extinction cross section there, as detected from ISO/SWS observations (Courtin et al. 2016). This instrument will also be necessary to determine the ice composition of the stratospheric polar clouds (e.g. Anderson et al. 2018).
- **An *in situ* Cosmorbitrap** (on a lander/drone) (Briois et al. 2016) will allow the detailed composition of the air, cloud, aerosols of the polar region through the descent and while the lander/drone will be operating near/at the surface.
- **An *in situ* Imager/spectral radiometer**: to observe solar aureole during the descent, derive the column opacity, the average aerosol and cloud particle size and their spectrum.
- **An *in situ* nephelometer/particle counter:** to determine the aerosol and cloud particles densities and size distribution.

**2.3 The lower atmosphere: clouds, weather and methane cycle**

*2.3.1 Current knowledge*

The deepest 80 km of Titan's atmosphere contains the deep stratosphere and the troposphere. This region, currently poorly known (mostly from its cloud activity and the *in situ* measurements performed by Huygens), is of particular interest because the interaction between surface and atmosphere occurs in the boundary layer (in the deepest 2 km) through convection. Some strong convective event can occur sporadically with cloud top reaching the tropopause at ≈40 km altitude (Griffith et al. 2005). Aerosols play an important role below 100 km, as they serve as condensation nuclei and are removed by sedimentation. The conditions of temperature and pressure on Titan allow the presence of a hydrological methane cycle, very similar to the Earth's water cycle. The weak solar flux reaching Titan's surface, and the generally dry conditions in the lower troposphere, lead to relatively rare tropospheric clouds (Griffith et al., 2005, 2006; Rodriguez et al., 2009, 2011; Turtle et al. 2009, 2011, 2018). Despite the scarcity of the observed tropospheric clouds, Cassini revealed their diversity including small patchy convective clouds, tropical storms associated to precipitations and stratospheric polar clouds. Large tropospheric clouds are thought to be composed of large methane droplets in ascending motions while thinner high altitude clouds are made of smaller ice crystals of photochemical by-products ($C_2H_6$, $C_2H_2$, HCN and other nitriles and hydrocarbons) in descending air (Rannou et al. 2006; Griffith et al. 2005, 2006; Barth and Toon 2006; Barth and Rafkin 2006; Hueso and Sanchez-Lavega 2006).

*2.3.2 Open questions*

$CH_4$, which plays a key role in the complex chemistry of Titan's atmosphere, comes from the surface and/or subsurface (**Section 3.5**). Monitoring the latitudinal distribution and seasonal evolution of tropospheric methane humidity may allow us to identify the $CH_4$ main evaporation sources (e.g. polar lakes, hypothetic tropical lakes or ground humidity). Cassini/CIRS observations also revealed unexplained large latitudinal variations of the methane abundance in Titan's stratosphere (Lellouch et al. 2014), which could be the result of methane injection from strong tropospheric convective events.

Even if several mechanisms have been proposed to explain the diversity, localization and seasonal evolution of observed tropospheric clouds, including planetary waves, global circulation, topography and boundary layer processes, a clear understanding of how clouds form, evolve and dissipate is missing. Their composition and the size of the cloud droplets are highly linked to their formation mechanism and are currently unknown. Getting a more complete climatology of Titan's clouds will provide strong constraints on the atmospheric circulation, the methane transport and the dominant mechanisms of cloud formation (global circulation, planetary waves, …). A key question related to the cloud formation and coverage is where and at which season it rains, and what are the precipitation rates. In particular, the dynamics, frequency and precipitation rates of convective methane storms are of prime interest to explain the formation of fluvial valley networks and equatorial dunes (Mitchell et al. 2012; Charnay et al., 2015) (see **Sections 3.1 and 3.2**).

The only direct measurements that we have about the wind speed in Titan's lower troposphere come from the Huygens probe descent, at a single epoch and a single location near the equator. A key question is to know the direction and speed distribution of surface winds as well as their seasonal evolution. In particular, atmospheric models predict that mean surface winds should be westward in the equatorial region (as trade winds on Earth), while dunes propagate eastward. The sand transport may be dominated by strong and rare eastward gusts produced by vertical mixing or by methane storms at the equinox (Tokano 2010; Charnay et al. 2015). In the same manner, large dust storms may have been detected by Cassini in the arid tropical regions of Titan (Rodriguez et al., 2018), but the strength of the surface winds able to generate them are still unknown. Another remaining question is to understand the atmospheric circulation over Titan's lakes and the timing and frequency of their wave activity.

Answering those open questions regarding the physic-chemical properties of Titan's low atmosphere remains a major goal for future missions to address. They are summarized here:
- **[1] What are the characteristics of the $CH_4$ cycle on Titan?**
- **[2] How do Titan's clouds form and evolve? What is their precipitation rate?**
- **[3] What is the wind regime near the surface?**

*2.3.3 Proposed instrumentation and mission concept to address the open questions*

**Understanding the methane cycle in the troposphere and its injection in the deep stratosphere (open question 1)** requires:
- **An orbiter far- and mid-IR spectrometer** and **submm spectrometer** to derive the $CH_4$ abundance independently of the temperature in the lower stratosphere all over the globe and especially above the northern lake region.
- **An *in situ* near-IR spectrometer** on a drone and/or a/several lander(s) to probe the methane abundance (humidity) below the tropopause and near the surface.

**Understanding cloud formation and humidity conditions in the troposphere requires (open question 2)**:
- **An orbiter radio occultation experiment** to determine the temperature profile from 150 km down to the surface.
- **An orbiter visible near-IR camera** to monitor cloud activity and precipitation signatures on surface.
- **An orbiter near-IR spectrometer** to derive the cloud composition and information on the $CH_4$ droplet size as well as other crystal sizes ($CH_4$, $C_2H_6$).
- **An *in situ* nephelometer/particle counter + camera/spectral radiometer + Cosmorbitrap** on a drone or on one/several lander(s) (with data acquisition during the descent) to determine the aerosol and cloud particles composition, their size distribution, and the cloud droplet phase (liquid, crystalline), and to look at the solar aureole to measure the scattered light and transmission through the atmosphere.

**Understanding the wind regime at/near the surface (open question 3)** requires *in situ* anemometer on a drone and/or on one/several landers localized in the northern lake region and at the equator. Some of the above mentioned *in situ* instruments (near-IR spectrometer, anemometer) will be carried by the Dragonfly mission that will study the dunes region with a planned arrival in 2034, i.e. during an expected "quiet" period, as the northern spring equinox will occur in 2039. An objective of our future Titan mission is specially to monitor the seasonal changes around equinoxes and if it arrives slightly before 2039, it will be highly complementary to the Dragonfly mission. The best science return from the two missions would be obtained if timing overlap could occur. The required measurements to answer the open questions cannot be performed from Earth ground-based/space-borne facilities.

# 3. Science Goal B: Titan's geology

Titan has diverse and strikingly familiar landscapes (mountains, rivers, seas, lakes, dunes, impacts…). Many of its surface morphologies are thought to originate from exogenic processes (Moore and Pappalardo 2011), first of all involving a complex and exotic climatology based primarily on the methane cycle, analogous to the hydrological cycle on Earth (Atreya et al. 2006). Nevertheless, endogenic processes (including a possible past and/or present tectonic and cryovolcanic activity) may also be at play (**Figure 3**). Even though the recent and remarkable progress accomplished so far in understanding Titan's thanks to the successful and long-lived Cassini-Huygens mission, numerous key issues regarding Titan geological history remain poorly constrained.

## 3.1 Aeolian features and processes

### 3.1.1 Current knowledge

Cassini observations revealed that dunes are Titan's dominant aeolian landform. Dunes, in particular, provide a powerful tool to investigate the sedimentary and climatic history of the arid and/or semi-arid environments likely to prevail at Titan's tropics. Therefore, the latitudinal distribution of these features is indicative of the different types of climates that Titan experiences or has experienced in the past.

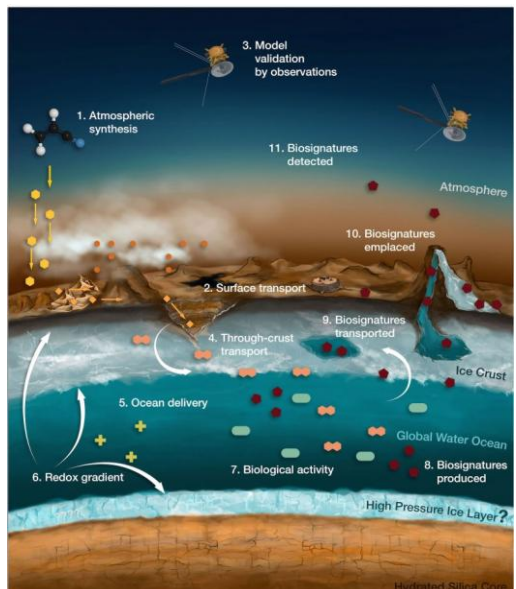

Figure 3: Molecules produced in Titan's atmosphere settle on the surface, where they participate in lateral transport by wind and liquid methane. If tectonic processes are also occurring, then these chemicals can be mixed downwards into the interior, undergoing further change and oxidation in an aqueous environment. (R. Lopes/JPL).

Titan's dunes are generally 1–2 km wide, spaced by 1–4 km and can be over 100 km long (Lorenz et al. 2006; Radebaugh et al. 2008). Limited estimation of heights from radarclinometry and altimetry suggests they are 60–120 m high (Neish et al. 2010; Mastrogiuseppe et al. 2014). By extrapolating the incomplete Cassini coverage of the surface it has been estimated that Titan's dunes cover up to ≈ 17% of the moon's total surface area (≈ $14.10^6$ km$^2$, 1.5 times the surface area of the Sahara desert on Earth - Rodriguez et al. 2014; Brossier et al. 2018). Dunes are mostly found within ±30° latitudes forming a nearly complete circum-Titan belt, at the notable – unexplained – exception of the Xanadu region (Barnes et al. 2015). The extent of the dunes indicates that sands have been generated on Titan in great volumes and transported by wind, and that processes have acted on the surface long enough to produce extensive and morphologically consistent landforms (Radebaugh 2013).

The interaction of Titan's dunes with topographic obstacles implies a general W-E transport of sand (Radebaugh et al. 2010). Dunes' size, morphology and relationship with underlying terrain, and their style of collection are similar to large, linear dunes in the Sahara, Arabia and Namibia (Lorenz et al. 2006; Radebaugh et al. 2008; Le Gall et al. 2011; 2012). Such dunes on Earth typically form under bimodal wind regime (Fryberger and Dean 1979; Tsoar 1983). A recent model calls on a dominant, slightly off-axis wind and a secondary wind causing sand flux down the dune long axis (Courrech du Pont et al. 2014; Lucas et al. 2014a). A fundamental challenge is to understand the eastward direction of sand transport (Lorenz et al. 2006; Radebaugh et al. 2010) as current climate models predict that low-latitude near-surface winds should generally blow to the west. Possibly, dunes reflect strong but infrequent eastward winds, either associated with vertical mixing in the atmosphere at equinox (Tokano 2010) or methane rainstorms (Charnay et al. 2015) leading to strong westerly gusts. Note that equatorial methane rainstorms may be associated with regional dust storms (Rodriguez et al. 2018). Additionally, convergence of the meridional transport predicted in climate models can explain why Titan's dunes are confined within ±30° latitudes, where sediment fluxes converge (Lucas et al. 2014a; Malaska et al. 2016).

Titan's dunes are not only consistently dark at Cassini's RADAR experiment wavelength (Elachi et al. 2004) but they are also some of the infrared-darkest materials observed by ISS (Porco et al. 2005), and have a low albedo and red slope in the near-infrared as seen by VIMS (Soderblom et al. 2007; Barnes et al. 2008; Clark et al. 2010; Rodriguez et al. 2014). This strongly indicates that the dunes are smooth at the 2.17 cm RADAR wavelength, homogeneous and primarily dominated by organic sand, presumably settling from the atmosphere and further processed at the surface (Soderblom et al. 2007; Barnes et al. 2008; Le Gall et al. 2011; Rodriguez et al. 2014; Bonnefoy et al. 2016; Brossier et al. 2018), making the dunes the largest carbon reservoir at Titan's surface (Lorenz et al. 2008a; Rodriguez et al. 2014). There is also evidence for sand-free interdunes (e.g. Barnes et al. 2008; Bonnefoy et al. 2016) implying that the dunes have been active on recent timescales.

In addition to the dunes, other aeolian processes, features and landforms on Titan's surface may have been identified by Cassini. These are possibly wind streaks (e.g. Porco et al. 2005; Lorenz et al. 2006; Malaska et al. 2016), yardangs (Paillou et al. 2016; Northrup et al. 2018), wind-carved ridges (Malaska et al. 2016) and dust storms (Rodriguez et al. 2018).

*3.1.2 Open questions*

The complex interplay between the hydrocarbon cycle, atmospheric dynamics and surface processes leading to the formation and dynamics of Titan's aeolian features is still far from being fully understood. The precise composition, grainsize and mechanical properties of Titan's sediment, its total volume, sources, transport dynamics and pathways at global scale still require further investigation.

Important open questions include:
- **What is the precise – not extrapolated – geographic distribution of Titan's aeolian landforms?**
- **What is the precise morphometry of the dunes and does it change with locations on Titan?**
- **What is the wind regime responsible to dunes' and other aeolian landforms' morphology and orientation?**
- **Are dunes and other aeolian landforms still active today?**
- **What are the sources and sinks of Titan's sand? Can we draw the pathways of sediment transport? Why are there no dunes in Xanadu?**
- **What is the composition, grain size, degree of cohesion and durability of the dune material?**

*3.1.3 Proposed key instrumentation and mission concept to address those questions*

Most of the open questions related to Titan's global distribution and properties of Titan's aeolian landforms (statistics on dunes' width, spacing, shape and height) can be addressed by a **Titan orbiter**, instrumented with a **multi-wavelength remote sensing package** (e.g. near-infrared in Titan's atmospheric windows' spectral range and microwaves) providing a decameter spatial resolution and a complete coverage by the end of the mission. This can be achieved either using a **SAR system**, or by using **an infrared camera** working in one of the methane spectral windows. An **imager at 2 µm** would provide the best tradeoff between signal to noise, atmospheric transparency and low aerosol scattering effects.

In order to study the accurate morphometry of a selection of aeolian landforms and the physico-chemical properties of Titan's sediment at the grain scale, a **mobile *in situ* probe** with **sampling** (e.g. a Cosmorbitrap-based High-resolution chemical analyzer), **imaging** and **spectral capabilities** is required.

**3.2 Fluvial features and processes**

*3.2.1 Current knowledge*

One of the most striking observations of Cassini-Huygens are Titan's channel networks (Collins 2005; Lorenz et al. 2008b; Lunine et al. 2008; Burr et al. 2006, 2009; Black et al. 2012; Burr et al. 2013). Similar to the Earth, Titan may experience complex climate-topography-geology interactions, involving surface runoff and subsurface flows. Valley networks are distributed at all latitudes (Lorenz et al. 2008b; Burr et al. 2009; Lopes et al. 2010; Langhans et al., 2012) and have a wide variety of morphologies: canyons at the poles (Poggiali et al. 2016), dendritic and rectilinear networks globally distributed (Burr et al. 2013), meandering feature in the south polar region (Malaska et al. 2011; Birch et al. 2018) and braided rivers at the equator (Lucas et al., 2014b). Near the poles, most of the fluvial networks are connected to empty or filled lacustrine features (cf. **Section 3.3**). Due to the hectometer resolution of the Cassini RADAR, only the largest valley networks on Titan could be studied. We therefore have a limited idea about the extent to which Titan's landscapes are dissected by fluvial networks. The one exception to this is the region where the Huygens lander descended and landed, where images, with a decameter to centimeter spatial resolution, showed a highly dissected network of dendritic valleys and rounded pebbles near its estuary (Tomasko et al. 2005), indicating that river networks at that scale may be frequent on Titan.

The mere presence of channelized flow conduits implies that flows of sufficient magnitude, from precipitation or groundwater, are able to erode Titan's surface either physically or chemically. However, using estimates for the initial topography and erodibility of the substrate, channels may be very inefficient agents of erosion on Titan (Black et al. 2012) or there may be a gravel lag deposit that inhibits erosion under Titan's current climate (Howard et al. 2016).

*3.2.2 Open questions*

Critical unknowns remain following the Cassini-Huygens mission. Better estimates for the physical and chemical properties of both the bedrock and the fluid (including frequency and magnitude of rain falls as a function of latitude) are needed to improve our understanding of role fluvial channels play in shaping Titan's surface (Burr et al. 2006; Cordier et al. 2017; Malaska et al. 2017; Richardson et al. 2018). Due to the limited coverage and spatial resolution of the Cassini remote sensing data, we have only limited constraints on the fluvial channel geographic distribution and morphologies. The question of the latitudinal (hence climatic) forcing on the dominant mechanisms of Titan's river network morphodynamics (which of them are alluvial channels formed in sediments or rather incision channels) is still open. Also, the possibility that Titan is dissected everywhere at the scale observed by Huygens, as it is the case on Earth, implies that Titan's landscape may be dominated by hillslopes. As hillslope processes control the sediment supply to rivers, it is important to clarify if slopes are made of consolidated material, or if they are covered with atmospheric sediment (i.e, granular media).

Many questions cannot be answered by the analysis of Cassini-Huygens or telescopic data:
- **What is the complete geographic distribution of river networks down to the decametric scale?**
- **How do river network morphology types vary with location?**
- **What are the processes at play forming the rivers (incision and/or dissolution) and what is the nature of the eroded material?**
- **What are the age and the current activity of the fluvial channels? How does this activity vary with latitudes?**

*3.2.3 Proposed key instrumentation and mission concept to address those questions*

Answers to these questions require observations with a resolution finer than the scale of fluvial dissection (10's of meters). A **long-living Titan orbiter** with a near-polar orbit and a **high-resolution remote sensing package** (down to decameter) will provide both the global coverage and a good repetitiveness needed (1) to build a consistent global map of the fluvial networks' distribution, (2) to provide a deeper look into their precise morphologies and possibly (3) to build digital elevation models of a variety of river networks by photogrammetry and/or radargrammetry. **Spectral capabilities are needed at both decameter spatial and high spectral (R>1000) resolutions** (**Figure 4b**) in order to help constraining the composition and texture of the eroded material, bedrock and transported sediment. The spectral identification of the surface component will always be limited by the strong atmospheric absorption, unless we develop a spectro-imager with a very high spectral resolution within the broader diagnostic methane windows (such as the 2, 2.7 and 5 µm windows)

A **mobile *in situ* probe** may be of great help to provide an unprecedented detailed view onto the morphology the river networks, the shape and size of the sediment and the composition of the involved materials (fluid, sediment, substrate and bedrock) with **remote sensing instruments** and **sampling capabilities** (including a Cosmorbitrap-based High-resolution chemical analyzer). The flexibility of an autonomous aerial drone would in addition provide the possibility to realize super high-resolution digital elevation models (down to the centimeter), allowing the analysis of the river dynamics down to the scale of boulders and pebbles. This could be done at high northern latitudes to complement the similar measurements that will be performed in the equatorial region by the Dragonfly quadcopter.

**3.3 Seas and lacustrine features and processes**

Titan's surface conditions (1.5 bar, 90 - 95 K) are close to the triple point of methane (and in the liquid stability zone of ethane), which allows standing liquid bodies to exist at the surface (Lunine and Atreya 2008). During the Cassini mission, Titan's surface has been unevenly mapped by all the imaging instruments (RADAR in SAR mode, VIMS spectrometer, and ISS) with various spatial resolutions (a few kilometers for ISS and VIMS, a few hundred of meters for the RADAR), extent (global coverage with ISS and VIMS, ≈50 % of the surface with the RADAR at 1500 m pixel resolution), and wavelengths (infrared and microwaves), looking for signs of these liquid bodies. Lacustrine features were first observed in 2004 in the infrared with the Cassini/ISS observation of Ontario Lacus (Porco et al. 2005), the largest lacustrine depression in Titan's south polar region. Titan's large seas and lacustrine features (lakes and topographic depressions) were then observed in 2006 when flying over the northern polar regions (Lopes et al. 2007; Stofan et al. 2007), as well as numerous fluvial features connected to the seas (cf. **Section 3.2**). Most of the liquids are currently located in the North, presumably as a result of orbital forcing (Aharonson et al. 2009; Hayes 2016).

*3.3.1 Current knowledge*

Titan's seas and lakes distinction is mainly based on their respective morphologies and size. Titan's large seas (Kraken Mare, Ligeia Mare, Punga Mare) are several 100s km-wide features with complex dissected shorelines shaped by rivers and drowned valleys from the nearby reliefs. Lacustrine features, regardless of their liquid-filling state, appear as closed rounded to irregularly lobate depressions, which are 10s to a few 100s km-wide and usually organized in clusters lying in flat areas (Hayes et al. 2017).

Lacustrine depressions sometimes display 100 m-high raised rims and wide ramparts. In a few cases, bathymetric profiles and rough composition could be determined by looking at double peaks in altimetry echoes over liquid-covered depressions (Mastrogiuseppe et al. 2014, 2019). Analyses of RADAR data provided a first estimate of the exposed volumes of liquids contained in the largest lakes and seas of Titan (70,000 km$^3$, Hayes et al. 2018), and show that lakes and seas consist in a relatively small methane reservoir as compared to the atmosphere, with varied compositions with respect to their locations and altitudes. Interestingly, the liquid-filled depressions at regional scale are at similar altitudes and are systematically lower than empty depressions, both being located at higher altitudes than the sea level (Hayes et al. 2017). This suggests the existence of a subsurface connection between the lakes such as an alkanofer surface replenishing low-altitude depressions (Cornet et al. 2012a; Hayes et al. 2017). An unknown portion of the liquid hydrocarbons could therefore well be stored in a permeable subsurface (as also suggested by models aiming at reproducing the stable polar liquids (Horvath et al. 2016)), and not considered in the organic inventory calculation on Titan. The accurate estimate of the liquid composition, topography and bathymetry of lacustrine features has a direct impact on constraining the amount of material removed and the total volume of liquid hydrocarbons stored in the seas and lakes.

Lakes and seas strongly differ in shape. The absence of well-developed fluvial networks at the 300 m-scale of the RADAR/SAR associated to the lacustrine features, in addition to the fact that they seem to grow by coalescence in areas hydraulically and topographically disconnected from the seas, suggests that a distinct scarp retreat process is responsible for the formation and evolution of the lacustrine features on Titan, for example raised rims and extensive ramparts (Cornet et al. 2012a; 2015; Hayes et al. 2016, 2017; Solomonidou et al. 2019). Among the hypothesis elaborated to explain the formation of the lacustrine features, the thermodynamical, geological and chemical contexts seem to favor the formation by karstic dissolution/evaporitic processes involving chemical dissolution/crystallization of solutes (soluble molecules) in solvents (in response to the rise or lowering of ground liquids).

Depending on its exact composition, the liquid phase is more or less stable under Titan's surface conditions, ethane and nitrogen giving more stability to the liquids (Luspay-Kuti et al. 2012, 2015). While the current search for lake changes within the Cassini dataset is still debated, Titan's surface exhibits hints of surface liquid changes at geological timescales. The identification of geological features that may have hold liquids in the recent or older past can help addressing this question. Currently Titan experiences a climate, which allows the long-term accumulation of polar liquids and brings liquids to the low latitudes only during torrential and sporadic events (Turtle et al. 2011b). Nonetheless, polar liquids are currently most seen in the North as a result of Saturn's current orbital configuration (Aharonson et al. 2009; Hayes 2016; Birch et al. 2018), while the South of Titan exhibit large catchment basins (Birch et al. 2018; Dhingra et al. 2018). Their total area is equivalent to that of the northern seas (Birch et al. 2018). A few lacustrine features may have been detected at lower latitudes (Griffith et al. 2012; Vixie et al. 2015), which indicates that the climate has evolved through time, potentially reversing the preferred location of liquid accumulation to the South in the past. By constraining the past climate of Titan, looking at the superficial record of the changes at the surface, one can reconstruct a climate history that will take part in constraining the methane cycle on Titan.

*3.3.2 Open questions*

Despite many observations spreading over the 13 years of the Cassini mission (from winter to summer in the northern hemisphere), a number of open questions remain regarding the methane cycle and the evolution of Titan's seas and lakes. In particular, we still do not have clear understanding of the formation scenarios of Titan's lakes, the precise composition of the liquid and substrate, the total volume of organics stored in the seas and lakes (and in a potential alkanofer), and how these organics are re-distributed over seasonal to geological timescales.

Uncertainties remain on the actual shape of these features due to the scarcity and accuracy of available topography data (currently provided at global scale and poor horizontal resolution by Corlies et al. 2017). Also, to date, only a few bathymetry profiles have been derived from RADAR altimetry data crossing liquid bodies. The bathymetry of Titan's largest sea, Kraken Mare, remains unknown. In the same manner, the exact composition of the solutes implied in dissolution/crystallization processes and of the solvent is to be determined. Radar altimetry data suggests the seas are primarily methane (Mastrogiuseppe et al. 2014), though VIMS observations detected ethane (Brown et al. 2008). Their mechanical properties, influencing the landscape evolution in response to mechanical/fluvial erosion and hillslope processes that can also contribute to some extent to the surface evolution, are also to be determined. The way and timescale, on which solids can accumulate over the surface to build the chemically eroded landscapes, has also to be constrained, notably by characterizing the thickness of the organic sedimentary layer being eroded.

The remaining major open questions on lacustrine features and processes on Titan are the following:
- **What are the shapes of the lacustrine features in the polar regions?**
- **What is the true distribution of sub-kilometer lakes and what does this tell us about lake formation?**
- **How much liquid is stored in the depressions?**
- **What are the exact compositions of the lakes and seas, how and why do they differ?**
- **By which geological processes do the lacustrine depressions and raised ramparts or rims form?**
- **What are the lake seasonal/short timescale changes?**
- **What is the true total inventory of organics in the polar areas?**

*3.3.3 Proposed key instrumentation and mission concept to address those questions*

A global topographic map at high vertical (10's of meters) and horizontal resolutions (a few hectometers) is required to address the major open questions regarding the total surface and sub-surface liquid organic inventory on Titan. At regional scale, at higher resolution, it will also help to constrain the formation of Titan's seas and lakes and connecting their distribution and properties with the present past climatic conditions.

A **long-living Titan orbiter** with a near-polar orbit will be required, including a **SAR system**, a **Ground Penetrating Radar system** and/or a **high-precision altimeter**. An *in situ* mobile/floating/submarine probe, including a **spectro-imager**, **electrical environment and meteorological packages**, and **sampling capabilities** (e.g. a Cosmorbitrap-based High-resolution chemical analyzer) would provide a fundamental support to the questions of lakes and substrates' topography and composition, and local geologic and climatic conditions.

## 3.4 Impact craters

*3.4.1 Current knowledge*

Cassini unveiled Titan's impact crater paucity. Cassini observations show that the resurface tendency of Titan, due to its very active atmosphere - similarly to Earth's, wiped out the majority of impact craters that hit the surface in the past, especially in the polar regions, leaving only approximately 60 potential craters as derived from Cassini RADAR and VIMS data (Wood et al. 2010; Buratti et al. 2012; Neish and Lorenz 2012). From these 60 potential impact craters, only 23 are labeled as 'certain' or 'nearly certain' (Lopes et al. 2019; Werynski et al. 2019). This scarcity of impacts features indicates that the surface is geologically young, $\approx$0.5-1 billion years old (Tobie et al. 2006; Neish and Lorenz 2012).

Cassini data show that Titan's craters are subject to extensive fluvial and aeolian erosion activity and infilling by sand (e.g. Le Mouélic et al. 2008; Soderblom et al. 2010; Neish et al. 2015; Brossier et al. 2018). In addition, Titan's craters are not uniformly distributed across the surface, where the Xanadu Regio area has the largest crater population and the poles have no crater at all (e.g. Neish and Lorenz 2014). There are a number of theories trying to explain this anomaly including inhibited formation from previous widespread seas (Neish et Lorenz 2014), burial due to heavy methane deposition or degradation due to fluvial erosion near the poles (Moore et al. 2014; Neish et al. 2016). An additional mystery about Titan's impact craters is their chemical composition. Cassini VIMS infrared data suggest that the very top layer of the impact craters could be dominated by atmospheric tholin-like material and that crater floors are rather constituted with water-ice rich materials from the upper lithosphere of the moon (Soderblom et al. 2010; Solomonidou et al. 2018; Brossier et al. 2018; Werynski et al. 2019). Also, craters, just after the impact, may be 'oases' where liquid water is in contact with organics for hundreds of years and thus the freshest may constitute interesting surface target to investigate the habitability of Titan.

*3.4.2 Open questions*

Cassini may have missed a number of small craters on Titan's surface due to fractional coverage and low spatial resolution. The cumulative crater-size frequency distribution available today is likely to be rather incomplete, and the relative ages of all of Titan's geologic units is still an open question. Also, craters provide invaluable windows into composition and mechanical properties of the crust, both of which are still largely unknown. Additionally, their present-day morphology gives key information on the strength of surface erosion by winds and rain falls.

Answering open questions regarding the impact craters on Titan remains a major goal for future missions to address:
- **What are the relative ages of all of Titan's geologic units?**
- **What is Titan's bedrock/crust composition?**
- **What are the erosion and degradation rates of craters? What do they reveal about Titan's past and present climatology? What is the reason for the difference in the crater population of Xanadu Regio from other regions on Titan, and in particular for the paucity of craters in Titan's polar regions?**

*3.4.3 Proposed key instrumentation and mission concept to address those questions*

At the end of the Cassini mission, only $\approx$45% of the surface has been imaged by SAR at 300-1500 m resolution, and 20% of the surface has been seen by VIMS with a resolution better than 10 km. Detailed geological investigations generally require at least under hectometer resolution, best would be decameter.

An **orbiter on Titan**, with high-resolution imaging capabilities (down to 10 meters) and overlapping use of instruments with infrared and microwave spectral capabilities would allow a systematic survey of impacts features, providing constraints on the processes that have shaped the moon, the age of the surface, and the composition of the surface and subsurface. A **high-resolution 2-µm imager** would provide the best tradeoff between S/N ratio and atmospheric transparency. A near-polar orbit with a sufficiently long mission will provide global surface coverage.

### 3.5 Interior-surface-atmosphere exchange processes

*3.5.1 Current knowledge*

The Cassini-Huygens mission collected three independent evidences for a **subsurface ocean** at Titan. The detection of electric disturbances by Huygens during its descent (Simões et al. 2007; Béghin et al. 2010), the obliquity value 3 times higher than what is expected for a solid interior (Stiles et al. 2010, Baland et al. 2011, 2014), and the temporal variations in Titan's gravitational potential (Iess et al. 2012; Mitri et al. 2014a; Durante et al. 2019) all indicate that Titan harbors a subsurface water ocean estimated at the depth of 50-100 km below the surface. The high amplitude of the tidal fluctuations estimated from the dynamic Love number, $k_2$ ($k_2 \approx 0.62 \pm 0.07$, Durante et al. 2019 suggests that it is a significantly denser layer than pure water, indicating a **high salt content** (Mitri et al. 2014a), which seems consistent with the detected electric signals (Béghin et al. 2010, 2012). Based on the phase diagram of aqueous solutions (e.g. Vance et al. 2018), Titan's ocean is probably sandwiched between the outer ice crust and a deep mantle composed of high-pressure (HP) ice phases that separate it from the silicate core, though solutions with a very dense ocean and no high pressure ice is compatible with the gravity field measurement by Cassini (Iess et al. 2012).

The Cassini-Huygens mission also provided crucial information on the long-wavelength topography (Zebker et al. 2009; Lorenz et al. 2013; Corlies et al. 2017) and low-degree gravity field of Titan, which put constraints on the **3D structure of the ice shell** (Choukroun and Sotin 2012; Hemingway et al. 2013; Lefevre et al. 2014; Mitri et al. 2014a; Kvorka et al. 2018). The long-wavelength topography is characterized by depression at the poles that could result either from accumulation of hydrocarbon clathrates (Choukroun and Sotin 2012) or from thinning of the ice shell (Lefèvre et al. 2014; Kvorka et al. 2018). The observed long-wavelength topography suggests that active exchange processes exist between the dynamical ocean and the ice shell, which may control the global geodynamics of Titan (Kvorka et al. 2018; Griffith et al. 2019). It has been also proposed that the organic-rich crust may locally or regionally be recycled to the underlying water ocean due to large impacts (Lunine et al. 2010; Zahnle et al. 2014) or Rayleigh-Taylor convective instabilities, even though there is still no observational constraint of potential recycling.

The **source of the atmospheric methane**, which is irreversibly destroyed by photochemistry on timescale of about 10-30 million years (Yung et al. 1984; Griffith et al. 2013), is one of the most puzzling mysteries on Titan. Outgassing by **cryovolcanism** has been proposed as a possible replenishment mechanism (e.g. Tobie et al. 2006; Lopes et al. 2007) and plausible cryovolcanic landforms (e.g. Sotra Patera, Mohini Sulcus, Tui Regio) have been identified based on their morphology (Wall et al. 2009; Lopes et al. 2007, 2013) or evidence of change at the surface (e.g. Nelson et al. 2009; Solomonidou et al. 2016). However, no thermal anomaly has been detected by Cassini in both IR and microwave domains, due to IR atmosphere opacity and low spatial resolution of microwave observations (Lopes et al. 2013).

The composition of the present-day atmosphere, as revealed by Cassini–Huygens, also contains **key information on its past evolution and the origin(s) of volatile compounds**. The detection of a significant amount of $^{40}$Ar (the decay product of $^{40}$K) by Cassini–Huygens (Niemann et al. 2005, 2010; Waite et al. 2005) indicated that a few per cent of the initial inventory was outgassed from the interior. The observed $^{40}$Ar abundance (Niemann et al. 2010) suggests an efficient leaching process and prolonged water-rock interactions (Tobie et al. 2012), which seems consistent with the apparent high salt content of the ocean (Mitri et al. 2014a). Cassini-Huygens and ground-based measurements provided isotopic ratios of H, C, N, and O in $N_2$, CO, $CH_4$, HCN and $C_2$ hydrocarbons at various altitudes in Titan's atmosphere (e.g. Mandt et al. 2012; Nixon et al. 2012). The measured $^{15}N/^{14}N$ ratio in $N_2$ is similar to the cometary values for $NH_3$ and HCN (Niemann et al. 2010), suggesting an origin as $NH_3$ trapped in the building blocks of Titan in conditions similar to comet formations (Mandt et al. 2014). The $^{13}C/^{12}C$ in methane allows little to no fractionation from possible primordial values, suggesting that methane has been present in the atmosphere for less than a billion years (Mandt et al. 2012). In the absence of a proper initial reference value, however, it is impossible to retrieve information on fractionation processes with confidence.

*3.5.2 Open questions*

Despite the unprecedented advances accomplished by Cassini-Huygens in our knowledge on Titan's internal structure, numerous fundamental questions are still unanswered, due to a lack of specific measurements or to large uncertainties. For instance, even if Titan is the icy moon for which we have the most complete and accurate set of geophysical data, in particular accurate determination of tidal Love number $k_2$, it is not possible to conclude with certainty about the density and depth of the ocean with the present data. The thickness of the high-pressure ice mantle, which control exchange processes between the silicate core and the ocean (Choblet et al. 2017; Kalousova et al. 2018), is totally unconstrained. Yet, such information is essential to assess the astrobiological potential of the ocean. In particular, a thin high-pressure ice shell at present would suggest that the ocean was in direct contact with the warm rocky core during most of its evolution, a situation comparable to Europa.

Another fundamental question is whether cryovolcanism takes place (or has taken place in a recent past) on Titan (e.g. Tobie et al. 2006; Lopes et al. 2007). Even if some surface features suggest cryovolcanic activity, their nature and origin are strongly debated (e.g. Moore and Pappalardo 2011).

Our understanding of the mechanism governing the formation of Titan's atmosphere and its possible replenishment through time relies on precise determination of (still lacking) isotopic ratios in N, H, C and O-bearing species in Titan's atmosphere and surface materials (organics, hydrates and ices).

To summarize, the remaining open questions are:
- **What are the depth, volume and composition of the subsurface liquid water ocean?**
- **Is Titan currently, or has it been in the past, cryovolcanically active?**
- **Are there chemical interactions between the ocean, the rock core and the organic-rich crust?**
- **How did Titan's atmosphere form and evolve with time in connection with the interior?**

*3.5.3 Proposed key instrumentation and mission concept to address those questions*

A combination of geophysical measurements from the orbit (**radio experiment, radar altimeter and sounder, radar imager, magnetometer & plasma package**) and from the ground (**seismometer, radio transponder and electric sensors and magnetometer**) is required to constrain the hydrosphere structure, and hence provide essential constraints on the ocean composition and the thickness both outer ice shell and high-pressure ice mantle (if any). This is a preliminary requirement for assessment of the astrobiological potential of its internal ocean. Surface sampling of erupting materials could reveal the salt composition of the ocean.

Measurement by a **high-precision mass spectrometer** of the ratio between radiogenic and non-radiogenic isotopes in noble gases (Ar, Ne, Kr, Xe) in Titan's atmosphere will also put fundamental constraints on whether and how water-rock interaction have occurred in Titan's interior. Comparison between isotopic ratio in N, H, C and O-bearing species in the atmosphere (gas and aerosols) and in collected surface materials, at different locations, will provide key informations on the volatile origin, if they were chemically reprocessed in the interior and/or at the surface.

Future missions could have on board a **high-resolution microwave radiometer** to look for thermal anomalies (hot spots) revealing possible still active cryovolcanism on Titan, which cannot be observed in optical and infrared domain due to atmospheric opacity. Microwaves also offer the prospect of sensing the shallow subsurface and thus may detect warmth from old lava flows i.e., lava flows which have cooled at the surface and thus have no more infrared emission signature but are still tens of K above ambient at depth. An **infrared spectrometer of higher spectral resolution**, especially at 1.59, 2, 2.7 and 5 μm, and overlapping capabilities with a **radar instrument** and **high- spatial resolution visible--infrared camera** would help on the identification of specific constituents (such as $NH_3$ or local enrichment in $CO_2$) and their spatial distribution that could only have internal origin, and thus function as an additional "smoking gun" for cryovolcanism on Titan. Detailed mapping of geomorphological features, composition, topography and subsurface sounding in regions of interest may also reveal areas where recycling processes have occurred.

## 4. Science Goal C: Titan's habitability

*4.1 Current knowledge*

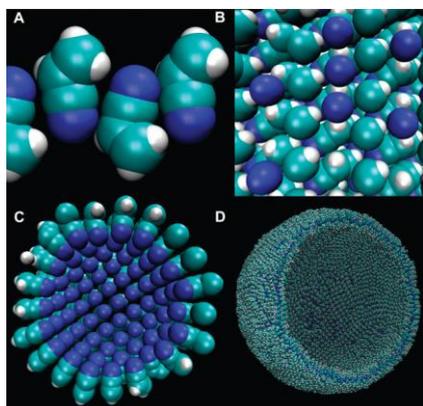

Figure 4: proposed structure of Titan "azotosome": a membrane that may be stable in liquid methane composed of acrylonitrile sub-units. Enclosed membranes are considered a vital prerequisite for life (Image: Cornell U./ Science Advances)

Habitable environments are defined as favoring the emergence and the development of life (Lammer et al. 2009). Habitability is based on the presence of a stable substrate, available energy, organic chemistry, and the potential for holding a liquid solvent. This definition identifies Titan as one of celestial bodies in the solar system with the highest potential for habitability. Titan harbors a complex organic chemistry producing a plethora of molecules and organic haze particles, and several sources of energy are available: solar radiation, solid body tides exerted by Saturn (Sohl et al. 2014), radiogenic energy production in the body core regions (Fortes 2012), and possible exothermic chemical reactions may be at work (Schulze-Makuch and Grinspoon 2005). Strikingly, two kind of solvents are simultaneously present on Titan: (1) **a dense, likely salt-rich, subsurface ocean with unconstrained fraction of ammonia** (Tobie et al. 2006; Iess et al. 2012; Béghin 2015), (2) **a mixture of simple hydrocarbons in liquid state**, forming a collection of seas and lakes in the polar regions (Stofan et al. 2007). These circumstances offer the possibility of the existence of **two distinct possible biospheres** between which fluxes of matter could be established, via geological processes like cryovolcanism. These chemical transfers are supported by the detection of $^{40}$Ar, the decay product of $^{40}$K initially contained in rocks, in the atmosphere (Niemann et al. 2010). Our knowledge about Titan's habitability, even if several interesting conceptual works have been published (McKay and Smith 2005; McKay 2016), remains very poor and essentially speculative. An on-site investigation is therefore invaluable for improving our views.

*4.2 Open questions*

Clearly, a potential "**aqueous Titan's biotope**", hidden below the icy crust, is far to be directly explorable. However, as already presented in **Section 3.5**, a series of geophysical measurements and chemical analysis of freshly erupted surface samples may provide crucial information on the ocean composition, on possible water-rock-organic interactions and internal heat sources. However, liquid water can be brought to the surface by cryovolcanism (Lopes et al. 2007) or cratering events (Artemieva and Lunine 2003). Once deposited on Titan's surface and in contact with liquid water, complex organic content produced by the atmosphere may lead to the production biologically important species such as amino acids and purines (Poch et al. 2012). Long-term chemical evolution is impossible to mimic experimentally in the laboratory. It is, therefore, crucial to be able to perform a detailed *in situ* chemical analysis of the **surface zones where cryovolcanism and impact ejecta (or melt sheets) are or have been present**, by using **direct sampling by drilling the near subsurface and/or performing spectroscopic observations**.

Hydrocarbons lakes and seas are also environments which could host a potential "**liquid hydrocarbons biosphere**". The exact chemical composition of these systems is also debated (Cordier et al. 2009; Luspay-Kuti et al. 2012; Tan et al. 2013), but it is generally well accepted that the liquid should be a mixture of three main species: methane, ethane and nitrogen, in variable proportions. A direct sampling is required to get a more solid picture of this composition. Regarding the role of these cryogenic solvents (the composition could be variable in time and space) is twofold: (1) liquids may interact with materials into which they are in contact, (2) the bulk volume may harbor a multitude of chemical processes. Probably the first question arising is the possible interaction between these cryogenic liquids and complex aerosols falling from the atmosphere. The transparency at the radar wavelength, of most of the lakes, indicates probably a low aerosol content (Mastrogiuseppe et al. 2019). However, the exact aerosol content of these lakes as well as the amount that could have sedimented at the lakes bottom still remain unconstrained. The presence of a floating film has been already proposed (Cordier and Carrasco 2019), the existence of such a deposit could be easily detected by a **Titan's lake lander**. The determination of its nature (monomolecular layer? thicker deposit made of aerosols?) is also important: the presence of a monomolecular layer could be the sign of the existence of a kind of "surfactant" which may have biological implications. Indeed, these class of molecules form micellas which could be the first stage of the formation of "vesicles", in which a specific a "proto-biochemistry" could appear. In a world with very little oxygen like Titan, analogs to terrestrial liposomes, entities called "azotosomes" have been already theoretically studied (Stevenson et al. 2015a,b) while their main compound, $C_2H_3CN$, has been recently detected in Titan's atmosphere (Palmer et al. 2017). This also reinforces the need of accurate chemical analysis of Titan's seas liquids, beyond the "simple" determination of $N_2+CH_4+C_2H_6$ mixing ratios.

Naturally, the interface ocean-atmosphere could reveal a large physico-chemical processes diversity. Recently, as an interpretation of the "Magic Islands" events (Hofgartner et al. 2014, 2016), bubbles streams coming from Ligeia Mare depths have been proposed (Cordier et al. 2017; Cordier and Liger-Belair 2018). This scenario could be confirmed by direct investigations, while it is important for exbiological activity since it implies the stability of the solvent and the sea/lake global mixing of material. **The exploration of these seas and lakes requires a multi-instruments approach going from global imaging, to more specific *in situ* chemical measurements** or indirect ones performed by specific instruments like those based of sound speed measurements (Cordier 2016).

**In the fringes of Titan's lakes, possible evaporites deposits** may exist, this is supported by infrared observations combined with radar imaging (Barnes et al., 2011) and also by numerical models (Cordier et al. 2013, 2016). According to these latter, due to its solubility properties, acetylene could be the dominant component of these evaporitic layers. In an astrobiological perspective, a massive presence of acetylene is not anecdotal. Indeed, at the surface of Titan the total solar radiation flux is only ≈0.1% of its terrestrial counterpart (Tomasko et al. 2005, 2008b). Titan's is then disfavoured compared to the Earth according to this aspect. This is the reason why, the hydrogenation of acetylene $C_2H_2$ (see equation below), if it occurs, has been proposed to ensure anabolism, according to $C_2H_2 + 3H_2 \rightarrow 2CH_4$, which releases a free energy of 334 kJ.mol$^{-1}$ of acetylene consumed, larger than the ≈40 kJ.mol$^{-1}$ required at least for methanogen growth on Earth (McKay 2016). The quantification of this chemical energy source available would require the measurement of the abundances of acetylene and molecular hydrogen (Strobel 2010) at the surface of Titan. If acetylene has been identified in the atmosphere (Coustenis et al. 2007); its presence, in solid state, in drybeds or lake beaches is not clear. **Drilling an evaporites field could bring numerous crucial constraints concerning the replenishment/evaporation cycle of Titan's lakes, together with interesting clues about a potential exotic microbiota**.

On Earth, chirality is a characteristic property of biogenic molecules. Then, detection of chiral molecules may be used as an indicator of some biological activity on Titan. In terrestrial context, it has been emphasized that living beings use chiral, stereo-chemically pure macromolecules (Plaxco and Allen 2002). These molecules show a noticeable circular dichroism in the terahertz domain which could be used as a general biosignature (Xu et al. 2003). To **determine chiral properties of samples collected** on the surface of Titan **requires the development of a dedicated instrument**, not yet available among those that had already flown or those currently under development.

The numerous remaining open questions regarding the potential habitability of Titan are summarized here:

- **What is the nature and quantity of material exchange between the subsurface ocean and the surface? In the past, did a form of life develop in the water pond, formed by cryovolcanism or bolide impacts?**
- **How the organic material falling from the atmosphere is physically/chemically processed at the surface? Does it exist some catalytic path for the hydrogenization of acetylene or other reactions?**
- **Does a layer of surfactant (or even thicker deposit) cover the surface of some lakes/maria?**
- **What is the nature of dissolved species in hydrocarbon lakes? Does this liquid environment harbor a chemical reactions network?**
- **Are the molecules present in lakes and evaporites deposits optically active? Can a kind of homochirality be exhibited?**

*4.3 Proposed key instrumentation and mission concept to address those questions*

A very high resolution mass spectrometer is needed, like the already mentioned **Cosmorbitrap** (Briois et al. 2016), should be used for low atmosphere composition measurement. An instrument of the same class is also needed for the analysis of liquid phases and solid surfaces (evaporitic terrains and crater soils), complemented by specific samplers for both phases. These **samplers should be a drill and an instrumented diving probe**, linked to a main "sea lander" (lake lander or amphibious drone) by a technical cable supplying power and commands, and collecting measurements signal. Concerning chirality determinations, a **chiral gas chromatograph** (Patil et al. 2018) will be perfectly suitable.

## 5. Mission concept

**5.1 Previous mission concepts for post-Cassini-Huygens exploration of Titan**

Future exploration of the Saturnian system with a focus on Titan has been considered for quite some time, almost since the first years of the Cassini-Huygens mission. The Titan explorer (Leary et al. 2008) and the Titan and Enceladus Mission (TandEM, Coustenis et al. 2009) concepts had been selected respectively by NASA and ESA for studies before they were merged into the joint large (Flagship) Titan and Saturn System Mission (TSSM) concept, which was extensively studied in 2008 (K. Reh and J. Lunine et al.–NASA, and C. Erd, J.-P. Lebreton and A. Coustenis et al.–ESA, TSSM NASA/ESA Joint Summary Report, 2009). TSSM aimed at an in-depth long-term exploration of Titan's atmospheric and surface environment with a dedicated orbiter, and *in situ* measurements in one of Titan's lakes (with a lake lander) and in the atmosphere with a montgolfière (hot air balloon).

TSSM was ranked second in the final decision by the agencies and was not considered for further study. It is still, however, one of the most ambitious mission concepts dedicated to Titan exploration to date, and has inspired several other proposed concepts, aborted or not selected, for smaller size missions and new instrumentations: **Titan Aerial Explorer** (TAE), a pressurised balloon (Hall et al. 2011); **Aerial Vehicle for *in situ* and Airborne Titan Reconnaissance** (AVIATR), an ASRG (Advanced Stirling Radioisotope Generator) powered airplane (Barnes et al. 2012); **Titan Mare Explorer** (TiME), a lake lander (Stofan et al. 2010); **Titan Lake Probe**, which included a submarine concept (Waite et al. 2010); **Journey to Enceladus and Titan** (JET), a single Saturn orbiter that would explore the plume of Enceladus and the atmosphere and surface of Titan (Sotin et al. 2011); a **seismic network** had been considered as part of the geophysical payload of such missions (Lorenz et al. 2009); **mission concepts with two elements: a Saturn-Titan orbiter and a Titan Balloon** (Tobie et al. 2014) and **a Saturn-Titan orbiter and a lake probe** (Mitri et al. 2014b).

Among the most recent proposals, Dragonfly (Lorenz et al. 2017), an extraordinary and inspiring mission concept, involving for the first time the use of an autonomous rotorcraft to explore *in situ* the surface and low atmosphere of Titan, has been selected by NASA in June 2019 as its 4[th] New Frontier mission. Dragonfly is scheduled for launch in 2026, arriving at Titan in 2034. The spacecraft will touch down in dune fields in the equatorial regions. From there Dragonfly will fly from location to location over a distance of 175 kilometers in two and a half years. Despite its unique ability to fly, Dragonfly would spend most of its time on Titan's surface making science measurements (sample surface material and measure with a mass spectrometer to identify the chemical components and processes producing biologically relevant compounds; measure bulk elemental surface composition with a neutron-activated gamma-ray spectrometer; monitor atmospheric and surface conditions, including diurnal and spatial variations, with meteorology and geophysical sensors; use imaging to characterize geologic features; perform seismic studies to detect subsurface activity and structure). In-flight measurements are also planned (contribute to atmospheric profiles below 4 km altitude, provide aerial images of surface geology, give context for surface measurements and scouting of sites of interest). Unable to use solar power under Titan's hazy atmosphere, Dragonfly will use a MMRTG (Multi-Mission Radioisotope Thermal Generator). Flight, data transmission, and most science operations will be planned during Titan's daytime hours (eight Earth days), giving the rotorcraft plenty of time during the Titan night to recharge. Dragonfly is a great step forward in the Solar System exploration history, technologically and scientifically speaking, pioneering the use of an extremely mobile and comprehensively instrumented *in situ* probe to investigate the atmosphere-surface-interior interactions of Titan. NASA and Dragonfly are paving the way to an ambitious, international program to explore the extreme complexity of Titan, hopefully in synergy and partnership with other agencies including ESA.

In May 2019, a mission concept (TOPS: Titan Orbiter/Polar Surveyor) was submitted to NASA for consideration as a major flagship study for the forthcoming Decadal Survey for Planetary Sciences, expected in 2021. This mission concept proposes an instrumented orbiter, and also 1-2 probes to land on Titan's largest lake(s): Kraken and/or Ligeia. If selected, TOPS mission would allow for significant synergy with the Dragonfly mission and involvement by international partners including ESA, with potentially contributions of instruments, subsystems or launch, subject to the appropriate bilateral agreements being negotiated. This would allow continuation of the highly successful scientific collaborations that made Cassini-Huygens so successful, and widen the participation of the international scientific community.

**5.2 An ESA L-class mission concept for the exploration of Titan**

In order to fulfil the totality of the key science goals presented in the present White Paper, we estimate that a L-class mission involving a Titan's orbiter and at least an *in situ* element that has sufficient mobility to probe the atmosphere **and the solid and possibly liquid surface (thus excluding balloon) is required**. Inspired by the experience of TSSM, such a mission, with international collaboration to support the total architecture and cost, will perfectly complement, and surpass, the exploration of Titan undertaken in the 2000s as part of the NASA-ESA-ASI Cassini-Huygens mission and reactivated today by the selection of the NASA Dragonfly mission concept and the TOPS proposal.

Titan's equinox would be the ideal season for observing tropical storms and their consequences for fluvial and aeolian features. It is also the best period to observe strong changes on the global atmospheric dynamics and its impact on the photochemical compounds distribution as well as on the thermal field. Equinoxes on Titan during the ESA Voyage 2050 period will be on 22 January 2039 (Northern spring equinox) and on 10 October 2054 (Northern autumn equinox). Having a mission with an orbiter planned to monitor the seasonal transition over the 2039 Northern spring equinox would take the extraordinary advantage to potentially overlap with the Dragonfly mission, complementing its scientific targets and even possibly acting as a transmission relay. We therefore advocate for a launch window in the early phase of the ESA Voyage 2050 cycle and even before. Considering an estimation of the cruise from Earth to Saturn to 7-8 years, a launch as early as 2031-2032 would be required. In the case of a partnership of ESA with NASA regarding the Dragonfly mission, our arrival at Titan should be as early as 2034. The clock is ticking! Note that any arrival outside those dates would still have, with the use of an orbiter and an *in situ* element, an outstanding scientific value, still answering fundamental open questions that remain about Titan's system that cannot be answered from Earth ground-based/space-borne facilities.

We propose that the **orbiter**, once arrived at Saturn, will be captured around Titan in an elliptical orbit followed by a few months of aero-braking. This aero-braking phase will enable the exploration of a poorly known, but chemically critical, part of the atmosphere (700-800 km), with *in situ* atmospheric sampling at altitudes much lower than possible with Cassini. Following the aero-braking phase, the orbiter will be placed into a near-polar elliptical orbit of lower eccentricity for the orbital science phase of a nominal duration of at least 4 years. Orbits' closest approach will be localized in the thermosphere to continue performing *in situ* mass spectra measurements for each orbit. It will also allow imagery and spectroscopy of the surface and atmosphere at all latitudes with high spatial and temporal resolutions, and repetitiveness.

While an orbiter would be of high value to provide global coverage, we are deeply convinced that the addition of at least an *in situ* probe is critical in terms of scientific complementarities with the orbiter for any next ambitious missions to Titan. In addition to the orbiter, which can serve as a transmission relay, we thus propose a mission scenario with **in situ element(s) to explore the Polar regions of Titan**: a lake lander, a drone and/or an air fleet of mini-drones. Titan's low gravity and dense atmosphere make it an ideal candidate for drone-based missions. The *in situ* probe(s) will be able to perform atmospheric measurements directly inside the polar vortex and image the surface during the descent phase.

- A **Titan lake lander** was already proposed in two mission candidates and one mission concept. Both mission candidates were close to be selected. Less mobile and more scientifically focused than a drone, it has the advantage to propose a mission element, and corresponding mission architecture, less complex and risky. A Titan lake lander can directly be built on ESA's heritage established with the Huygens probe and does not require critical new technological developments. The lander would preferentially float and passively drift for days at a lake's surface (with long-lived batteries), possibly including the additional capability of plunging. Such a probe will directly sample and observe the lake's liquid properties (temperature, viscosity, permittivity, composition, wave/current activity and tidal deformation), as well as its geological context (shorelines and surroundings' composition and morphology, depth) and local meteorological conditions, which cannot be done from orbit, or only with extreme difficulty, larger uncertainties and lower resolution.

- A **Titan heavy drone** would propose an incredibly better alternative to the immobility (or very low mobility) of a lake lander and **has our preference**. Also proposed by the AVIATR consortium, the idea to use an autonomous flying drone to explore Titan's low atmosphere and surface is an original concept and thus requires the resolution of numerous challenges and the development of numerous innovative technologies. The feasibility study of the AVIATR unmanned aircraft and the recent selection by NASA of Dragonfly, a Mars' rover-sized quadcopter, give however the promises of very near-future operational drones for planetary exploration to all agencies and to the entire scientific community. In perfect complement to Dragonfly time, area and topics of exploration (equatorial geology and meteorology at Northern winter), a Titan flying drone in the North Polar regions, arriving soon before Northern spring equinox, would provide an extraordinary tool for exploring the complex geology and meteorology of the lakes' district in a context of changing seasons. Equipped with MMRTGs, the 'lake' drones would be able to fly from one lake to another, and land close to their shorelines, and directly sample and measure, with an unprecedented precision, regional variations in composition and geology, but also in local wind, humidity, pressure and precipitations, and how the climatology interacts with the liquid bodies. If we add the ability to fly up to 10 km altitude and also land and float a few minutes on lakes (and sample their liquids), we have the ideal and most comprehensive *in situ* element concept to explore Titan's polar region geology and low atmosphere.

- A **Titan air fleet of mini-drones** would be an even more revolutionary concept than a unique drone, but would guarantee at the same time a larger range of action and the concomitant study of multiple targets, without being possibly less risky than a unique large drone. Again, we propose this concept to explore the Polar regions of Titan in order to perfectly complement the scientific goals of the Dragonfly mission. The mini-drones (e.g. mono-copter cubes of a few inches in size), that could be inspired from the "Mars Helicopter Scout", a mini-copter that will fly on Mars with the Mars 2020 rover, would be of course less instrumented than a large drone, but could nevertheless fulfil a wealth of observations and measurements by including a well focused miniaturized payload (cameras, meteorological and electric environment package, ...). The air fleet could be composed of a few mini-drones that could be released by the orbiter one by one by sequence, depending on the location of the targeted lakes (one at the South Pole to explore Ontario Lacus and the rest at the North Pole – a unique opportunity to *in situ* investigate two poles at the same time). A platform could also be deployed near the air fleet at North Pole that can serve as a communication relay with the mini-drones and the orbiter, and as a possible recharging station for a greater longevity.

All those *in situ* probe concepts are thought to be highly complementary, in terms of targets, measurements and/or achievable resolution, to the orbiter, but also to all available observatories that will operate at (or close to) the time of the ESA Voyage 2050 program (Dragonfly, JWST, ALMA, large Earth-based telescopes with Adaptive Optics).

**Strawman instrument payload: Table 1** presents a tentative payload that would address the required measurements for the science goals A, B, C. The proposed instruments will benefit from the heritage of successful missions such as Cassini-Huygens, Rosetta, Venus and Mars Express, as well as new missions currently under study (such as JUICE, ExoMars, ...).

*Table 1: Tentative instrument payload to address the three mission goals A, B and C.*

| Titan Orbiter | Titan probe (lander, drone(s)) *Mini-drones possible payload is considered in blue* |
|---|---|
| 1. High spatial resolution Imager (2, 2.7, 5-6 µm) and high spatial and spectral resolution (R>1000) near-IR Spectrometer (0.85-6 µm) **[A,B,C]** | 1. Visual Imaging System (two narrow angle stereo cameras and one wide angle camera) **[A,B,C]** |
| 2. Radar active and passive imager **[B,C]** | 2. Imaging Spectrometer (1-5.6 µm) **[A,B,C]** |
| 3. Penetrating Radar and Altimeter (> 20 MHz) **[B,C]** | 3. Atmospheric Structure Instrument and Meteorological Package **[A]** |
| 4. Mid- to Far-Infrared Spectrometer (5-1000 µm) **[A,B,C]** | 4. Electric Environment and Surface Science Package (including a penetrator and a drill) **[A,B,C]** |
| 5. Cosmortbitrap – High Resolution Mass spectrometer (up to 10000 amu) **[A,C]** | 5. Radar sounder (> 150 MHz) **[B,C]** |
| 6. Icy grain and organic Dust analyzer **[A,B,C]** | 6. Cosmorbitrap – Gas Chromatograph Mass Spectrometer (1-600 amu) **[A,B,C]** |
| 7. Plasma suite **[A,B,C]** | 7. Radio science using spacecraft telecom system **[A,B,C]** |
| 8. Magnetometer **[A,B,C]** | 8. Magnetometer **[A,B,C]** |
| 9. Radio Science Experiment **[A,B,C]** | 9. Neutron-activated gamma-ray spectrometer **[B,C]** |
| 10. Sub-Millimetre Heterodyne Receiver **[A,B,C]** | 10. Seismometer **[B,C]** |
| 11. UV Spectrometer **[A,B,C]** | |

We propose in particular that one of the new *in situ* key instruments to be carried on the orbiter and the *in situ* element(s) will be a Cosmorbitrap, which will acquire mass spectra with unequalled mass resolution. This instrument has a mass resolution 100 times higher than the mass spectrometer that will be on board Dragonfly (a Curiosity/SAM or ExoMars/ MOMA type spectrometer). **Figure 5-left** shows the interest of high mass resolution to determine without ambiguity the composition of a gas, solid or liquid. The Cosmorbitrap is currently TRL 5. It has a mass of 8 kg and needs 50 W of power, and therefore could be included in a drone/lander payload. Significant new insights into Titan's atmosphere and surface composition will also come from the use of near-infrared hyperspectral imager with a major increase in spectral range (specially above 5 µm where numerous organics have diagnostic absorption bands), and spectral and spatial resolution. This is exemplified in **Figure 5-right**, where the ethane 2-µm absorption band diagnostic structure starts being revealed thanks to a spectral resolution 10 to 30x times the resolution of Cassini/VIMS.

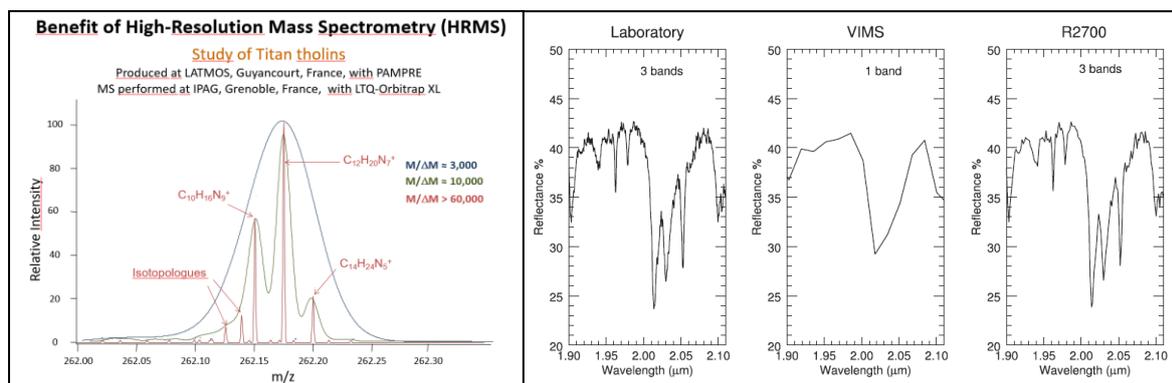

Figure 5: (Left) Mass spectrum of aerosol analogues (tholins) acquired at different mass resolution. Cosmorbitrap will have a resolution M/DM > 60,000 (the Cassini/INMS mass spectrometer had a resolution of 500). (Right) Comparison of a laboratory spectrum of liquid ethane at 2.0 µm, with the same absorption viewed at Cassini/VIMS spectral resolution (≈100) and at spectral resolution 2700. The diagnostic triple band at 2 µm shows up only at high spectral resolution (>1000). Laboratory spectra from the Arkansas Center for Space and Planetary Sciences.

**Critical issues and technological developments:** Beyond Jupiter, solar power is inefficient and radioisotope power sources are the only alternative. In the TSSM concept, MMRTGs or ASRGs using $^{238}$Pu were considered and were to be provided by NASA. Within Europe the radioisotope $^{241}$Am is considered a feasible alternative to $^{238}$Pu and can provide a heat source for small-scale radioisotope thermoelectric generators (RTGs) and radioisotope heating units (RHUs) (Sarsfield et al. 2012), albeit at higher mass. $^{241}$Am exists in an isotopically pure state within stored civil plutonium at reprocessing sites within the UK and France – about 1000 kg of $^{241}$Am exist in the civil PuO$_2$ stockpile of the UK and France. A study is underway to design a process that will chemically separate $^{241}$Am (Sarsfield et al. 2012). The development of $^{241}$Am-based RTGs is under consideration by ESA and should be available at high TRL before the proposed Voyage 2050 launch windows.

Drones have never been concretely considered by ESA for planetary exploration. However, their technology seems now mature for spatialization and is of greatest value for the exploration of remote places such as planetary atmospheres and surfaces. Following the impulsion given by NASA and the selection of Dragonfly, we advocate that ESA conducts technical analyses expressly dedicated to the feasibility study and technological developments of planetary flying drones. A detailed comparison between the different approaches (one heavily instrumented drone, an air fleet of mini-drones or a lake lander) will be needed to determine the best option for *in situ* exploration of Titan's atmosphere and surface.

Instrumenting a drone or the lake lander's heat shield with a geophysical and meteorological package (a seismometer and possibly other instruments, such as a drill, a penetrometer, an electrical environment package, an anemometer, a pressure sensor...) can also be considered. Instrumenting the probes' heat shield was already considered by TSSM. Such options would require further study to evaluate their feasibility and utility. Finally, supports from national agencies will be essential in developing the new generation of highly capable instrumentations, as well as in pursuing experimental and modelling efforts initiated with Cassini-Huygens, in order to be ready for this next rendezvous with Titan.

**Possibility for downgrading the mission class:** Mission scenarios with different budget envelopes could be, by increasing order of mission class: a **M-class** a mission concept with only an *in situ* element (lake lander, drone, a few mini-drones), but missing all the key science questions related to global processes, a **L-class** a mission concept including a Titan's orbiter and small, focused *in situ* element(s) (lake probe, a few mini-drones), missing key questions at regional scale, and a **L+-class** (with international collaboration to support the total architecture and cost) a mission concept including a Titan's orbiter and at least one ambitious *in situ* element (a flying, 'amphibious'/floating drone), allowing to address all the fundamental questions summarized in this document.

Should the mission concept be descoped, one could also consider the participation of ESA with another space agency to provide a (partial) support for an orbiter including key instruments (Cosmorbitrap, submm spectrometer, near-IR and visible spectro-imager, and small plasma and magnetosphere *in situ* analysis suites) or an *in situ* probe for Titan exploration, in synergy with the incoming NASA mission(s) to Titan. In that sense, for instance, ESA could propose an orbiter to go with Dragonfly from the beginning of its mission and have the quadcopter as the *in situ* element. As well, the TOPS concept being selected by NASA for further feasibility study, ESA could participate by providing a second *in situ* element or the lake lander, or participate to their conception.